\documentclass[12pt]{article}
\usepackage{amsfonts}
\usepackage{amssymb}
\usepackage{amsthm}
\usepackage{amsmath}
\usepackage{bm}
\usepackage{graphicx}
\begin{document}

\renewcommand{\theequation}{\thesection.\arabic{equation}}

\title{The averaging of non-local Hamiltonian structures
in Whitham's method.}

\author{Andrei Ya. Maltsev}

\date{
\centerline{L.D.Landau Institute for Theoretical Physics, 117940}
\centerline{ul. Kosygina 2, Moscow, maltsev@itp.ac.ru} 
\centerline{and}
\centerline{SISSA-ISAS, Via Beirut 2-4 - 34014 Trieste , ITALY}
\centerline{maltsev@sissa.it} 
}

\maketitle

\begin{abstract}
 We consider the $m$-phase Whitham's averaging method
and propose a procedure of ``averaging'' of non-local
Hamiltonian structures. The procedure is based on the
existence of a sufficient number of local commuting
integrals of a system and gives a Poisson bracket of 
Ferapontov type for the Whitham's system. The method 
can be considered as a generalization of the
Dubrovin-Novikov procedure for the local
field-theoretical brackets.
\end{abstract}

\centerline{\bf{Introduction.}}

We consider the averaging of non-local Hamiltonian
structures in Whitham's averaging method. 
As it is well known, the Whitham's method permits 
to obtain equations on the ``slow'' modulated 
parameters of exact periodic or quasi-periodic 
solutions of systems of partial differential equations and it
was pointed out by Whitham (\cite{whith}) that these equations
can be written in the Lagrangian form if the initial system 
possesses a local Lagrangian structure. The Lagrangian formalism 
for the Whitham's system is given in this case by the ``averaging'' 
of a local Lagrangian function, defined for the initial system, 
on the corresponding space of (quasi)-periodic solutions.
Some basic questions concerning Whitham's method can be found in 
\cite{whith,luke,ffm,dn1,krichev1,krichev2,dn2,krichev3,dn3,dm,
dobr1,dobr2,DobrKrichever,DobrMinenkov}.

 B.A. Dubrovin and S.P. Novikov investigated also the question
of the conservation of local field-theoretical Hamiltonian
structures in Whitham's method and suggested a procedure
of ``averaging'' of a local field-theoretical Poisson bracket, 
giving a Poisson bracket of Hydrodynamic type for the 
Whitham system (\cite{dn1,dn2,dn3}, see also \cite{novmal}).

 The Jacobi identity for the averaged bracket and the 
invariance of the Dubrovin-Novikov procedure was
proved by the author in \cite{engam} 
(see also \cite{umn1}) using the 
Dirac restriction procedure of the initial bracket on the 
subspace of quasi-periodic ``$m$-phase'' solutions of the initial
system. The connection between the procedure of Dubrovin and
Novikov and the procedure of averaging of the Lagrangian function
in the case when the initial local Hamiltonian structure just
follows from the local Lagrangian one was also studied in
\cite{malpav}. Some extension of the averaging of ``local'' 
Hamiltonian structures for the case of discrete systems is also 
presented in \cite{umn2}.

 In the present work we consider the Poisson brackets having
a non-local part 

\begin{equation}
\label{operat1}
\{\varphi^{i}(x), \varphi^{j}(y)\} \,\,\, = \,\,\, \sum_{k\geq 0}
B^{ij}_{k}(\varphi,\varphi_{x},\dots) \, \delta^{(k)}(x-y) 
\,\, + \,\, \sum_{k\geq 0} \,\, 
e_{k} \, S^{i}_{(k)}(\varphi,\varphi_{x},\dots)
\,\, \nu (x-y) \, S^{j}_{(k)}(\varphi,\varphi_{y},\dots) 
\end{equation}
where $\, e_{k} = \pm 1$, $\, \nu (x-y) = - \nu (y-x)$,
$\, \partial_{x} \nu (x-y) = \delta (x-y)$ and both sums contain
a finite number of terms depending on a finite number of
derivatives of $\varphi$ with respect to $x$.

 Let us also point out here that the brackets (\ref{operat1}) 
usually appear in the theory the so-called ``integrable'' 
hierarchies (see \cite{Sokolov,EnOrRub,PhysD}), connected with 
the method of the inverse scattering problem.

 The most general form of the non-local Hamiltonian operators
(\ref{operat1}) containing only $\, \delta^{\prime}(X-Y)$ and
$\, \delta (X-Y)$ in the local part and the quasi-linear fluxes 
$\, S^{\nu}_{(k)\lambda}(U) \, U^{\lambda}_{X}$ of ``hydrodynamic'' 
type in the non-local one 
$$\{ U^{\nu}(X), U^{\mu}(Y) \} \,\,\, = \,\,\, g^{\nu\mu}(U) \,
\delta^{\prime}(X-Y) \,\, + \,\, 
b^{\nu\mu}_{\lambda}(U) U^{\lambda}_{X} \, \delta (X-Y) 
\,\,\, + $$
$$\,\,\, + \sum_{k \geq 0} \, e_{k} \, S^{\nu}_{(k)\lambda}(U) \,
U^{\lambda}_{X} \,\,\, \nu(X-Y) \,\, 
S^{\mu}_{(k)\delta}(U) \, U^{\delta}_{Y}
\quad , \quad \quad \quad
1\leq \nu,\mu,\lambda,\delta\leq N $$
was suggested by E.V.Ferapontov in
\cite{fer1} as a generalization of the bracket
introduced in \cite{mohfer1} and is usually called 
a weakly non-local Poisson bracket of Hydrodynamic type. 
We will discuss here a possibility of ``averaging'' of the 
brackets (\ref{operat1}) in the Whitham's method to obtain 
the bracket of such ``Hydrodynamic type'' for the Whitham 
system.
 
 As was shown by E.V. Ferapontov, the Hamiltonian
operators of this type reveal a beautiful 
differential-geometrical structure following from the
Jacobi identity of the bracket (\cite{fer1,fer2,fer3,fer4}).
(In particular they can be obtained as the Dirac restriction of
local differential-geometrical Poisson brackets on a submanifold
with flat normal connection (\cite{fer2}).)

 The first example of the non-local bracket (of Mokhov-Ferapontov
type, see \cite{mohfer1}) for the Whitham's system for NS equation
in the one-phase case was constructed by M.V.Pavlov in
\cite{pavlov} from a nice differential-geometrical consideration.
After that there was set the question about a possibility of
constructing of nonlocal Hamiltonian structures for
Whitham's system from the structures (\ref{operat1}) for the
initial one.
As was mentioned above, the Hamiltonian operators (\ref{operat1}) 
exist for many ``integrable'' systems like KdV and in the 
paper \cite{fer4} (see also \cite{alekspav}) 
there was a discussion of the possibility of 
averaging of the non-local operators for KdV equation using
the local bi-Hamiltonian structure and the recursion operator
for two averaged local Poisson brackets. The corresponding 
calculations for the $m$-phase periodic solutions of KdV were
made by V.L. Alekseev in \cite{alekseev} .

 Here we propose a general construction for the averaging
of operators (\ref{operat1}) in the Whitham's method which
gives a generalization of the Dubrovin-Novikov procedure for 
the case of presence of non-local terms in the bracket.
Our procedure does not require a local bi-Hamiltonian
structure and can be used in the general situation. 
Like in the procedure of Dubrovin and Novikov, we require 
here the existence of a sufficient number of commuting local
integrals, generating local flows according to 
(\ref{operat1}), and we also impose some conditions of 
``regularity'' of the full family of $m$-phase solutions as 
in the local case (see \cite{engam}).

\vspace{1cm}

\section{Some general properties of the non-local brackets.}
\setcounter{equation}{0}

  Let us consider a non-local 1-dimensional Hamiltonian 
structure of the type:

$$\{\varphi^{i}(x),\varphi^{j}(y)\} \,\,\, = \,\,\, 
\sum_{k \geq 0} \, B^{ij}_{k} (\varphi, \varphi_{x}, \dots) \, 
\delta^{(k)} (x-y) \,\,\, + $$
\begin{equation}
\label{nonlocbr}
+ \,\,\,\sum_{k \geq 0} \,
{\tilde S}^{i}_{(k)} (\varphi, \varphi_{x}, \dots) \,\, 
\nu (x-y) \,
{\tilde T}^{j}_{(k)} (\varphi, \varphi_{y}, \dots) 
\quad , \quad \quad \quad 1 \leq i,j \leq n
\end{equation}
where we have finite numbers of terms in both sums
depending on a finite number of derivatives of $\varphi$
with respect to $x$. 

 We will call a local translationally invariant Hamiltonian 
function a functional of the form:

\begin{equation}
\label{hamilt}
H [\varphi] \,\,\, = \,\,\, \int 
{\cal P}_{H} (\varphi, \varphi_{x}, \dots) \, dx
\end{equation}

 Here $\, \nu (x-y)$ is the skew-symmetric function
\begin{equation}
\label{nyxy}
\nu (x-y) \,\,\, = \,\,\, {1 \over 2} \,\, 
{\rm sgn} \, (x-y) \quad , \quad \quad
D_{x} \, \nu (x-y) \,\,\, = \,\,\, \delta (x-y) \,\,\, , 
\end{equation}
and $\, \delta^{(k)} (x-y)$ is the $k$-th derivative of the 
delta-function with respect to $x$.

  We assume here that the bracket (\ref{nonlocbr}) is 
written in the ``irreducible'' form, which means that the number 
of terms in the second sum is the minimal possible and the sets
$\, \{ {\tilde {\bf S}}_{(k)} \} $ 
and $\, \{ {\tilde {\bf T}}_{(k)} \} $ 
represent two linearly independent
sets of vector-functions of the variables 
$\, (\varphi, \varphi_{x}, \dots)$.
From the skew-symmetry of the bracket (\ref{nonlocbr}) it 
follows then that the sets of $\, {\tilde {\bf S}}_{(k)}$ and 
$\, {\tilde {\bf T}}_{(k)}$ define actually the same linear
space in the space of functions and it can be easily seen that 
the bracket (\ref{nonlocbr}) can be represented in the 
``canonical'' form
\begin{equation}
\label{canform}
\{\varphi^{i}(x) , \varphi^{j}(y)\} \,\,\, = \,\,\, 
\sum_{k \geq 0} \, B^{ij}_{k} (\varphi, \varphi_{x}, \dots) \,
\delta^{(k)}(x-y) \,\, + \,\,
\sum_{k \geq 0} \, e_{k} \, 
S^{i}_{(k)}(\varphi, \varphi_{x}, \dots) \,\, \nu (x-y) \,
S^{j}_{(k)}(\varphi, \varphi_{y}, \dots)
\end{equation}
where $\, e_{k} = \pm 1$ .

Indeed, since the sets $\, \{ {\tilde {\bf S}}_{(k)} \} $ and 
$\, \{ {\tilde {\bf T}}_{(k)} \} $ span the same linear space
we have just one finite-dimensional space,
generated by fluxes (vector fields)
$$\varphi^{i}_{\tau_{k}} \,\,\, = \,\,\, {\tilde S}^{i}_{(k)} 
(\varphi, \varphi_{x}, \dots)$$
and a symmetric (view the skew-symmetry of the bracket
and the function $\, \nu (x-y)$) finite-dimensional constant 
2-form, which describes their couplings in the non-local part 
of (\ref{nonlocbr}). So, we can write it in the canonical form
according to its signature after some linear transformation
of the flows $\, {\tilde {\bf S}}_{(k)}$ and
$\, {\tilde {\bf T}}_{(k)}$ with constant coefficients.

 We should also define in every case the functional space
where we consider the action of the Hamiltonian operator
(\ref{canform}) and this can depend on a concrete
situation. The most natural thing is to consider the 
functional space $\varphi(x)$ and the algebra of functionals 
$I[\varphi]$, such that their variational derivatives, 
multiplied by the flows 
$\, S_{(k)}(\varphi,\varphi_{x},\dots)$,
give us rapidly decreasing functions as
$|x| \rightarrow \infty$. Below we will use the functionals
of the type
$$\int \sum_{p=1}^{n} \,\, \varphi^{p}(x) \, q_{p}(x) \, dx 
\quad  , $$
where $q_{p}(x)$ are arbitrary smooth functions with compact
supports, to examine the properties of the bracket 
(\ref{canform}). For all the other functionals used
in the considerations we will assume that they have a
compatible with the bracket (\ref{canform}) form in the
sense discussed above.

 We will assume here for simplicity that the functions
$\, B^{ij}_{k}$ and $\, S^{i}_{(k)} $ represent
analytic functions of their arguments 
(maybe in some open region of the values of
$\, (\varphi, \varphi_{x}, \dots)$).

\vspace{1cm}

 We will construct here a procedure, which gives us 
a bracket of Ferapontov type (\cite{fer1}-\cite{fer4})
$$\{ U^{\nu}(X), U^{\mu}(Y) \} \,\,\, = \,\,\, 
g^{\nu\mu}(U) \,\, \delta^{\prime}(X-Y) \,\, + \,\, 
b^{\nu\mu}_{\lambda}(U) \, U^{\lambda}_{X} \,\,
\delta (X-Y) \,\,\, + $$
\begin{equation}
\label{ferbr}
+ \,\,\, \sum_{k \geq 0} \, e_{k} \, S^{\nu}_{(k)\lambda}(U) \,
U^{\lambda}_{X} \,\,\, \nu(X-Y) \,\, 
S^{\mu}_{(k)\delta}(U) \, U^{\delta}_{Y}
\quad , \quad \quad \quad 
1\leq \nu,\mu,\lambda,\delta\leq N
\end{equation}
from the initial bracket (\ref{canform})
after the averaging on an appropriate family of exact 
$m$-phase solutions of a local system, which is supposed
to be Hamiltonian with respect to the bracket (\ref{canform}) 
with a local Hamiltonian function $H$. 

 So, we consider here the Whitham's method for the local fluxes 
(if they exist)
\begin{equation}
\label{locsys}
\varphi^{i}_{t} \,\,\, =  \,\,\, 
Q^{i} (\varphi,\varphi_{x},\dots) \,\,\, ,
\end{equation}
generated by the Hamiltonian functions (\ref{hamilt}) 
in the non-local Hamiltonian structure (\ref{canform}).

 Certainly, this situation can arise in general only for special 
Hamiltonian functions, so all the considerations here appeal
as a rule to the ``integrable systems'' like KdV, NS, etc.,
where we have a lot of such functionals. 

 Let us now formulate some general theorem about the non-local
part of the bracket (\ref{canform}).

\vspace{1cm}

{\bf Theorem 1.1}

{\it 
Suppose we have a non-local Hamiltonian operator 
written in the ``canonical'' form (\ref{canform}),
where all $\, B^{ij}_{k}$ and $\, S^{i}_{(k)} $
represent analytic functions of 
$\, (\varphi, \varphi_{x}, \dots)$ 
in some open region of their values.

Then, for the same region of the values of
$\, (\varphi, \varphi_{x}, \dots)$:

1) The flows 

\begin{equation}
\label{sflows}
{\dot \varphi}^{i} \,\,\, = \,\,\, 
S^{i}_{(k)} (\varphi,\varphi_{x},\dots)
\end{equation}
commute with each other.

2) Any of the flows (\ref{sflows}) conserves the Hamiltonian
structure (\ref{canform}) }

\vspace{0.5cm}

{\it Proof.}

 Let us consider the functional

\begin{equation}
\label{ifq}
\int \sum_{p=1}^{n} \, \varphi^{p}(x) \, q_{p}(x) \, dx
\end{equation} 
for some $q_{p}(x)$ with compact supports and consider
the Hamiltonian flow $\xi^{i}(x)$, generated by (\ref{ifq})
according to (\ref{canform}), i.e.

$$\xi^{i}(x) \,\,\, = \,\,\, \sum_{k\geq 0} \, 
B^{ij}_{k}(\varphi,\varphi_{x},\dots) \,\,
{d^{k} \over dx^{k}} \, q_{j}(x) \,\,\, + $$   
\begin{equation}
\label{ksi}
+ \,\,\, {1 \over 2} \, \sum_{k\geq 0} \,\, e_{k} \, 
S^{i}_{(k)}(\varphi,\varphi_{x},\dots) \left[
\int_{-\infty}^{x} S^{j}_{(k)}(\varphi,\varphi_{z},\dots) \,
q_{j}(z) \, dz \,\, - \,\, \int_{x}^{\infty} 
S^{j}_{(k)}(\varphi,\varphi_{z},\dots) \,
q_{j}(z) \, dz \right] 
\end{equation}
(we assume summation over the repeated indices).

For the Hamiltonian flow $\xi^{i}(x)$ we should have:

\begin{equation}
\label{nulder}
\left[{\cal L}_{\xi} {\hat J}\right]^{ij}(x,y) 
\,\,\, \equiv \,\,\, 0
\end{equation}
where ${\hat J}$ is the Hamiltonian operator (\ref{canform})
and ${\cal L}_{\xi}$ is the Lie-derivative, given by the
expression:

$$\left[{\cal L}_{\xi} {\hat J}\right]^{ij}(x,y) 
\,\,\, = \,\,\,
\int \xi^{s}(z) \, {\delta \over \delta \varphi^{s}(z)}
\, J^{ij}(x,y) \,\, dz \,\,\, - $$ 
$$- \,\,\, \int J^{sj}(z,y) \, 
{\delta \over \delta \varphi^{s}(z)} \, \xi^{i}(x) \, dz 
\,\,\, - \,\,\,
\int J^{is}(x,z) \, {\delta \over \delta \varphi^{s}(z)}
\, \xi^{j}(y) \,\, dz $$

 Let us now consider the relation (\ref{nulder}) for $x$ and $y$
larger than any $z$ from the supports of $q_{p}(z)$. Then we will
have
$${\xi}^{i}(x) \,\,\, = \,\,\, \sum_{k\geq 0} \, e_{k} \,
S^{i}_{(k)}(\varphi,\varphi_{x},\dots)
\left[ {1 \over 2} \int_{-\infty}^{\infty} 
S^{p}_{(k)}(\varphi,\varphi_{w},\dots) \,
q_{p}(w) \, dw \right] $$
and

$$\left[{\cal L}_{\xi} {\hat J}\right]^{ij}(x,y) \,\,\, = \,\,\,
\sum_{k\geq 0} {\dot B}^{ij}_{k}(\varphi,\varphi_{x},\dots) \,
\delta^{(k)}(x-y) \,\,\, + $$
$$+ \,\,\, \sum_{k\geq 0} \, e_{k} \, 
{\dot S}^{i}_{(k)}(\varphi,\varphi_{x},\dots) \,\, \nu (x-y) \,
S^{j}_{(k)}(\varphi,\varphi_{y},\dots) \,\,\, + $$
$$+ \,\,\, \sum_{k\geq 0} \, e_{k} \,
S^{i}_{(k)}(\varphi,\varphi_{x},\dots) \,\, \nu (x-y) \,
{\dot S}^{j}_{(k)}(\varphi,\varphi_{y},\dots) \,\,\, - $$
$$- \,\, \sum_{k\geq 0} \, (-1)^{k} \, {d^{k} \over dy^{k}} \left(
B^{sj}_{k}(\varphi,\varphi_{y},\dots) \sum_{k^{\prime}\geq 0}
e_{k^{\prime}} 
{\delta S^{i}_{(k^{\prime})}(\varphi,\varphi_{x},\dots)
\over \delta \varphi^{s}(y)} \right) \, {1 \over 2} 
\int_{-\infty}^{\infty} 
S^{p}_{(k^{\prime})}(\varphi,\varphi_{w},\dots) \, q_{p}(w) \, dw
\,\,\,  - $$
$$- \,\,\,\,\, \sum_{k\geq 0} \, B^{is}_{k}(\varphi,\varphi_{x},\dots)
\, {d^{k} \over dx^{k}} \left( \sum_{k^{\prime}\geq 0}
e_{k^{\prime}}
{\delta S^{j}_{(k^{\prime})}(\varphi,\varphi_{y},\dots)
\over \delta \varphi^{s}(x)} \right) \, {1 \over 2}
\int_{-\infty}^{\infty}
S^{p}_{(k^{\prime})}(\varphi,\varphi_{w},\dots) \, q_{p}(w) \, dw
\,\,\,\,\, - $$
\begin{multline*}
- \,\,\, \int dz \, \sum_{k\geq 0} \, e_{k} \, 
S^{s}_{(k)}(\varphi,\varphi_{z},\dots) \,\, \nu (z-y) \,
S^{j}_{(k)}(\varphi,\varphi_{y},\dots) \, \times   \\
\times \, \sum_{k^{\prime}\geq 0} \, e_{k^{\prime}} \,
{\delta \over \delta \varphi^{s}(z)} \left(
S^{i}_{(k^{\prime})}(\varphi,\varphi_{x},\dots) \left[
{1 \over 2} \int_{-\infty}^{\infty} 
S^{p}_{(k^{\prime})}(\varphi,\varphi_{w},\dots) 
\, q_{p}(w) \, dw \right] \right) \,\,\, - 
\end{multline*}
\begin{multline*}
- \,\, \int dz \, \sum_{k\geq 0} \, e_{k} \,
S^{i}_{(k)}(\varphi,\varphi_{x},\dots) \,\, \nu (x-z) \,
S^{s}_{(k)}(\varphi,\varphi_{z},\dots) \, \times   \\
\times \, \sum_{k^{\prime}\geq 0} \, e_{k^{\prime}} \,
{\delta \over \delta \varphi^{s}(z)} \left(
S^{j}_{(k^{\prime})}(\varphi,\varphi_{y},\dots) \left[
{1 \over 2} \int_{-\infty}^{\infty}
S^{p}_{(k^{\prime})}(\varphi,\varphi_{w},\dots) \,
q_{p}(w) \, dw \right] \right)
\end{multline*}
where ${\dot B}^{ij}_{k}(\varphi,\varphi_{x},\dots)$ and 
${\dot S}^{i}_{(k)}(\varphi,\varphi_{x},\dots)$
are the derivatives of these functions with respect to
the flow
\begin{equation}
\label{potok}
{\varphi}^{i}_{t} \,\,\, = \,\,\, \sum_{k\geq 0} \, e_{k} \,
S^{i}_{(k)}(\varphi,\varphi_{x},\dots)
\left[ {1 \over 2} \int_{-\infty}^{\infty} 
S^{p}_{(k)}(\varphi,\varphi_{w},\dots) \,
q_{p}(w) \, dw \right] 
\end{equation}

 Here we also used that $\, x,y \, > \, Supp \,\, q_{p} \, $ 
when omitted the variational derivatives with respect to
$\varphi^{s}(x)$ and $\varphi^{s}(y)$ of the non-local expressions
containing the convolutions with $q_{p}(w)$ (the 4-th and the 5-th
terms).

 So we have

$$ 0 \,\,\, \equiv \,\,\, 
\left[{\cal L}_{\xi} {\hat J}\right]^{ij}(x,y) \,\,\, =$$
$$= \,\,\, \sum_{k\geq 0} \,  
{\dot B}^{ij}_{k}(\varphi,\varphi_{x},\dots) \, \delta^{(k)}(x-y) 
\,\, + \,\, \sum_{k\geq 0} \, e_{k} \,
{\dot S}^{i}_{(k)}(\varphi,\varphi_{x},\dots) \,\, \nu (x-y) \,
S^{j}_{(k)}(\varphi,\varphi_{y},\dots) \,\,\, +$$
$$+ \,\,\, \sum_{k\geq 0} \, e_{k} \, 
S^{i}_{(k)}(\varphi,\varphi_{x},\dots) \,\, \nu (x-y) \,
{\dot S}^{j}_{(k)}(\varphi,\varphi_{y},\dots) \,\,\, -$$
$$- \,\,\sum_{k\geq 0} \, (-1)^{k} \, {d^{k} \over dy^{k}} \left(
B^{sj}_{k}(\varphi,\varphi_{y},\dots) \sum_{k^{\prime}\geq 0}
e_{k^{\prime}} 
{\delta S^{i}_{(k^{\prime})}(\varphi,\varphi_{x},\dots)  
\over \delta \varphi^{s}(y)} \right) \, {1 \over 2}
\int_{-\infty}^{\infty} 
S^{p}_{(k^{\prime})}(\varphi,\varphi_{w},\dots) 
\, q_{p}(w) \, dw \,\,\, -$$
$$- \,\,\,\,\, 
\sum_{k\geq 0} \, B^{is}_{k}(\varphi,\varphi_{x},\dots)
\, {d^{k} \over dx^{k}} \left(
\sum_{k^{\prime}\geq 0} e_{k^{\prime}}
{\delta S^{j}_{(k^{\prime})}(\varphi,\varphi_{y},\dots)
\over \delta \varphi^{s}(x)} \right) \, {1 \over 2}
\int_{-\infty}^{\infty}
S^{p}_{(k^{\prime})}(\varphi,\varphi_{w},\dots) 
\, q_{p}(w) \, dw \,\,\, -$$
\begin{multline*}
- \,\,\, \int dz \, \sum_{k\geq 0} \, e_{k} \,
S^{s}_{(k)}(\varphi,\varphi_{z},\dots) \,\, \nu (z-y) \,
S^{j}_{(k)}(\varphi,\varphi_{y},\dots) \, \times   \\
\times \, \sum_{k^{\prime}\geq 0} \, e_{k^{\prime}} \,
{\delta S^{i}_{(k^{\prime})}(\varphi,\varphi_{x},\dots)
\over \delta \varphi^{s}(z)} \left[ {1 \over 2}
\int_{-\infty}^{\infty}
S^{p}_{(k^{\prime})}(\varphi,\varphi_{w},\dots) 
\, q_{p}(w) \, dw \right] \,\,\, -
\end{multline*}
\begin{multline*}
- \,\,\, \int dz \, \sum_{k\geq 0} \, e_{k} \,
S^{i}_{(k)}(\varphi,\varphi_{x},\dots) \,\, \nu (x-z) \,
S^{s}_{(k)}(\varphi,\varphi_{z},\dots) \, \times  \\ 
\times \, \sum_{k^{\prime}\geq 0} \, e_{k^{\prime}} \,
{\delta S^{j}_{(k^{\prime})}(\varphi,\varphi_{y},\dots)
\over \delta \varphi^{s}(z)} \left[ {1 \over 2}
\int_{-\infty}^{\infty}
S^{p}_{(k^{\prime})}(\varphi,\varphi_{w},\dots)
\, q_{p}(w) \, dw \right] \,\,\, -
\end{multline*}
\begin{multline*}
- \,\,\, \int dz \, \sum_{k\geq 0} \, e_{k} \,
S^{s}_{(k)}(\varphi,\varphi_{z},\dots) \,\, \nu (z-y) \,
S^{j}_{(k)}(\varphi,\varphi_{y},\dots) \, \times   \\
\times \, \sum_{k^{\prime}\geq 0} \, e_{k^{\prime}} \,
S^{i}_{(k^{\prime})}(\varphi,\varphi_{x},\dots) 
\,\, {1 \over 2}
\left[ {\delta \over \delta \varphi^{s}(z)} 
\int_{-\infty}^{\infty}
S^{p}_{(k^{\prime})}(\varphi,\varphi_{w},\dots)
\, q_{p}(w) \, dw \right] \,\,\, -
\end{multline*}
\begin{multline*}
- \,\,\, \int dz \, \sum_{k\geq 0} \, e_{k} \,
S^{i}_{(k)}(\varphi,\varphi_{x},\dots) \,\, \nu (x-z) \,
S^{s}_{(k)}(\varphi,\varphi_{z},\dots) \, \times   \\
\times \, \sum_{k^{\prime}\geq 0} \, e_{k^{\prime}} \,
S^{j}_{(k^{\prime})}(\varphi,\varphi_{y},\dots) 
\,\, {1 \over 2}
\left[ {\delta \over \delta \varphi^{s}(z)}
\int_{-\infty}^{\infty}
S^{p}_{(k^{\prime})}(\varphi,\varphi_{w},\dots)
\, q_{p}(w) \, dw \right] \,\,\, \equiv 
\end{multline*}

\vspace{1cm}

$$\equiv \,\,\, \sum_{k\geq 0} \, e_{k} \, 
\left[ {1 \over 2} \int_{-\infty}^{\infty}
S^{p}_{(k)}(\varphi,\varphi_{w},\dots)
q_{p}(w) dw \right] \cdot 
\left[{\cal L}_{k} {\hat J}\right]^{ij}(x,y) 
\,\,\, + $$
\begin{multline*}
+ \,\,\, \sum_{k,k^{\prime}\geq 0} \, e_{k} \,
\, e_{k^{\prime}} \,\, 
S^{i}_{(k^{\prime})}(\varphi,\varphi_{x},\dots) \,
S^{j}_{(k)}(\varphi,\varphi_{y},\dots) \, \times  \\
\times \, {1 \over 2} \int \left(  
S^{s}_{(k)}(\varphi,\varphi_{z},\dots) \,\,
\nu (y-z) \, {\delta \over \delta \varphi^{s}(z)} \left[
\int_{-\infty}^{\infty}
S^{p}_{(k^{\prime})}(\varphi,\varphi_{w},\dots)
\, q_{p}(w) \, dw \right] \right. \,\,\, -  \\
- \,\,\, \left.  
S^{s}_{(k^{\prime})}(\varphi,\varphi_{z},\dots) \,\,
\nu (x-z) \, {\delta \over \delta \varphi^{s}(z)} \left[
\int_{-\infty}^{\infty}
S^{p}_{(k)}(\varphi,\varphi_{w},\dots)
\, q_{p}(w) \, dw \right] \right) \, dz 
\end{multline*}
where $\left[{\cal L}_{k} {\hat J}\right]^{ij}(x,y)$ 
represent the Lie derivatives of $\, {\hat J}$ with 
respect to the flows (\ref{sflows})

$${\dot \varphi}^{i} \, = \,  
S^{i}_{(k)}(\varphi,\varphi_{x},\dots) $$

 Let us use again our condition 
$\, x,y \, > \, Supp \,\, q_{p} \, $
and rewrite the above identity in the form
$$0 \,\,\, \equiv \,\,\, \sum_{k\geq 0} \, e_{k} \, 
\left[ {1 \over 2} \int_{-\infty}^{\infty}
S^{p}_{(k)}(\varphi,\varphi_{w},\dots) \,
q_{p}(w) \, dw \right] \cdot 
\left[{\cal L}_{k} {\hat J}\right]^{ij}(x,y) 
\,\,\, + $$
\begin{multline*}
+ \,\,\, \sum_{k,k^{\prime}\geq 0} \, e_{k} \,
\, e_{k^{\prime}} \,\, 
S^{i}_{(k^{\prime})}(\varphi,\varphi_{x},\dots) \,
S^{j}_{(k)}(\varphi,\varphi_{y},\dots) \, \times  \\
\times \, {1 \over 4} \int \left(  
S^{s}_{(k)}(\varphi,\varphi_{z},\dots) \,
{\delta \over \delta \varphi^{s}(z)} \left[
\int_{-\infty}^{\infty}
S^{p}_{(k^{\prime})}(\varphi,\varphi_{w},\dots)
\, q_{p}(w) \, dw \right] \right. \,\,\, -  \\
- \,\,\, \left.  
S^{s}_{(k^{\prime})}(\varphi,\varphi_{z},\dots) \,
{\delta \over \delta \varphi^{s}(z)} \left[
\int_{-\infty}^{\infty}
S^{p}_{(k)}(\varphi,\varphi_{w},\dots)
\, q_{p}(w) \, dw \right] \right) \, dz 
\end{multline*}
 
 Using the standard expression for the variational derivative 
and the integration by parts we obtain that this identity can 
be written also in the form
$$0 \,\,\, \equiv \,\,\, \sum_{k\geq 0} \, e_{k} \, 
\left[ {1 \over 2} \int_{-\infty}^{\infty}
S^{p}_{(k)}(\varphi,\varphi_{w},\dots) \,
q_{p}(w) \, dw \right] \cdot 
\left[{\cal L}_{k} {\hat J}\right]^{ij}(x,y) 
\,\,\, + $$
$$+ \,\,\, \sum_{k,k^{\prime}\geq 0} \, e_{k} \, 
e_{k^{\prime}} \, 
S^{i}_{(k^{\prime})}(\varphi,\varphi_{x},\dots) \,
S^{j}_{(k)}(\varphi,\varphi_{y},\dots) \,\, {1 \over 4}
\int_{-\infty}^{\infty} \, q_{p}(z) \left[ S_{(k)} ,
S_{(k^{\prime})}\right]^{p}(z) \,\, dz $$
where $[ S_{(k)},S_{(k^{\prime})}]$ 
is the commutator of the flows (\ref{sflows}), or
\begin{equation}
\label{nulder1}
0 \,\,\, \equiv \,\,\, \sum_{k\geq 0} \, e_{k} \, 
\left[ {1 \over 2} \int_{-\infty}^{\infty}
S^{p}_{(k)}(\varphi,\varphi_{w},\dots) \,
q_{p}(w) \, dw \right] \cdot 
\left[{\cal L}_{k} {\hat J}\right]^{ij}(x,y) 
\,\,\, + 
\end{equation}
\begin{multline*}
+ \,\,\, \sum_{k > k^{\prime} \geq 0} 
e_{k} \, e_{k^{\prime}}
\left( S^{i}_{(k^{\prime})}(\varphi,\varphi_{x},\dots) 
\, S^{j}_{(k)}(\varphi,\varphi_{y},\dots) \,\, - \,\,
S^{i}_{(k)}(\varphi,\varphi_{x},\dots) \,
S^{j}_{(k^{\prime})}(\varphi,\varphi_{y},\dots) 
\right) \times  \\
\times \,\, {1 \over 4}
\int_{-\infty}^{\infty} \, q_{p}(z) \left[ S_{(k)} ,
S_{(k^{\prime})}\right]^{p}(z) \,\, dz 
\end{multline*}
for any $\, q_{p}(z)$, such that 
$\, x,y \, > \, Supp \,\, q_{p}(z) \, $.

 As can be easily seen, the last term in (\ref{nulder1}) 
represents the non-local part of 
$\left[{\cal L}_{\xi} {\hat J}\right]^{ij}(x,y) $,
which does not contain the function $\, \nu (x-y) $.
The first term in (\ref{nulder1}) also contains a
non-local part, however, this part contains the function
$\, \nu (x-y) $. It is not difficult to see that this
non-local part can in general be written in the 
``canonical'' form 
\begin{equation}
\label{SecondTerm}
\sum_{k\geq 0} \, e_{k} \, 
\left[ {1 \over 2} \int_{-\infty}^{\infty}
S^{p}_{(k)}(\varphi,\varphi_{w},\dots) \,
q_{p}(w) \, dw \right] \,\,\,
\sum_{s=1}^{Q} \, e^{\prime}_{s} \, 
A^{i}_{(s)} (\varphi,\varphi_{x},\dots) \,\,
\nu (x-y) \, A^{j}_{(s)} (\varphi,\varphi_{y},\dots)
\end{equation}
$(e^{\prime}_{s} = \pm 1) \, $, where the functions 
$\, {\bf A}_{(s)} (\varphi,\varphi_{x},\dots)$ 
represent some linearly independent set of
analytic vector-functions of 
$\, (\varphi,\varphi_{x},\dots) $. Let us prove now,
that from the identity (\ref{nulder1}) it follows 
actually that both the expressions (\ref{SecondTerm})
and
\begin{multline}
\label{FirstTerm}
\sum_{k > k^{\prime} \geq 0} 
e_{k} \, e_{k^{\prime}}
\left( S^{i}_{(k^{\prime})}(\varphi,\varphi_{x},\dots) 
\, S^{j}_{(k)}(\varphi,\varphi_{y},\dots) \,\, - \,\,
S^{i}_{(k)}(\varphi,\varphi_{x},\dots) \,
S^{j}_{(k^{\prime})}(\varphi,\varphi_{y},\dots) 
\right) \times  \\
\times \,\, {1 \over 4}
\int_{-\infty}^{\infty} \, q_{p}(z) \left[ S_{(k)} ,
S_{(k^{\prime})}\right]^{p}(z) \,\, dz 
\end{multline}
should be identically equal to zero.

 Let us fix some value of $x$ and consider the interval 
$\, I \, = \, [ x - \Delta , \, x + \Delta ] $,
such that  \linebreak 
$\,\, x - \Delta, \, x + \Delta \, > \, Supp \,\, 
q_{p} (z) $. It is
not difficult to see that for a linearly independent
set of analytic functions 
$\, {\bf A}_{(s)} (\varphi,\varphi_{y},\dots) \, $
we can find an everywhere dense set $\, {\cal S} $
of analytic on the interval
$\, y \, \in \, [ x - \Delta , \, x + \Delta ] \, $
functions $\, \varphi (y) \, $ (and infinitely smooth
on the whole numerical axis), such that the functions
$\, {\bf A}_{(s)} (\varphi,\varphi_{y},\dots) \, $
give a linearly independent set of analytic functions
of $y$ on the interval $\, I \, $ for any
$\, \varphi (y) \, \in \, {\cal S} \, $. It is easy to see
also, that for any $\, \varphi (y) \, \in \, {\cal S} \, $
we can find a set of analytic functions 
$\, \kappa_{i} (y) \, $ on the interval $\, I \, $,
such that the functions
$$a_{(s)} (y) \,\,\, = \,\,\, 
A^{i}_{(s)} (\varphi,\varphi_{y},\dots) \,\,
\kappa_{i} (y) $$
still give a set of linearly independent analytic
functions on $\, I \, $.

 According to Peano (\cite{Peano}), we can claim that
there exists a point $\, y_{0} \, \in \, I \, $ such that
the Wronskian
$$W (y_{0}) \,\,\, = \,\,\, 
\begin{vmatrix}
a_{(1)} (y_{0})  &  a_{(1) y} (y_{0})  &  \dots  &
a_{(1) (Q-1)y} (y_{0})  \\
\vdots  &  \vdots  &  \vdots  &  \vdots  \\
a_{(Q)} (y_{0})  &  a_{(Q) y} (y_{0})  &  \dots  &
a_{(Q) (Q-1)y} (y_{0})
\end{vmatrix}  $$
is different from $0$ at the point $y_{0}$. It is not
difficult to see also, that we can assume actually
that $\, y_{0} \, = \, x \, $, $\, $ so we put
$\,\, W (x) \, \neq \, 0 \, $.

 Let us introduce now infinitely smooth functions
$\,\, \zeta^{0} (y) , \, \dots , \, \zeta^{Q-1} (y) \, $
having the following properties:

\vspace{3mm}

1) All $\, \zeta^{l} (y) \, $ are identically equal to
zero outside the interval $\, I \, $;

\vspace{1mm}

2) All $\, \zeta^{l} (y) \, $ and all their derivatives
$\,\, \zeta^{l}_{sy} (y) \, $, $\,\, s \geq 0 \, $,
are equal to zero at the point $\, y = x \, $;

\vspace{1mm}

3)

$$\int_{-\infty}^{+\infty} \, y^{s} \,\, \zeta^{l} (y)
\,\, d y \,\,\, = \,\,\, 0  \quad , \quad \quad
\quad 0 \, \leq \, s \, < \, l \,\,\, , $$
$$\int_{-\infty}^{+\infty} \, y^{l} \,\, \zeta^{l} (y)
\,\, d y \,\,\, = \,\,\, l ! $$

\vspace{1mm}

4) The functions 
$\,\, {\hat \zeta}^{(l)} (y) \,\, = \,\, \nu (x-y) \,
\zeta^{l} (y) \,\, $ satisfy the relations
$$\int_{-\infty}^{+\infty} \, y^{s} \,\, 
{\hat \zeta}^{l} (y) \,\, d y \,\,\, = \,\,\, 0  
\quad , \quad \quad
\quad 0 \, \leq \, s \, \leq \, l $$

\vspace{3mm}

\begin{figure}[t]
\begin{center}
\includegraphics[width=0.9\linewidth]{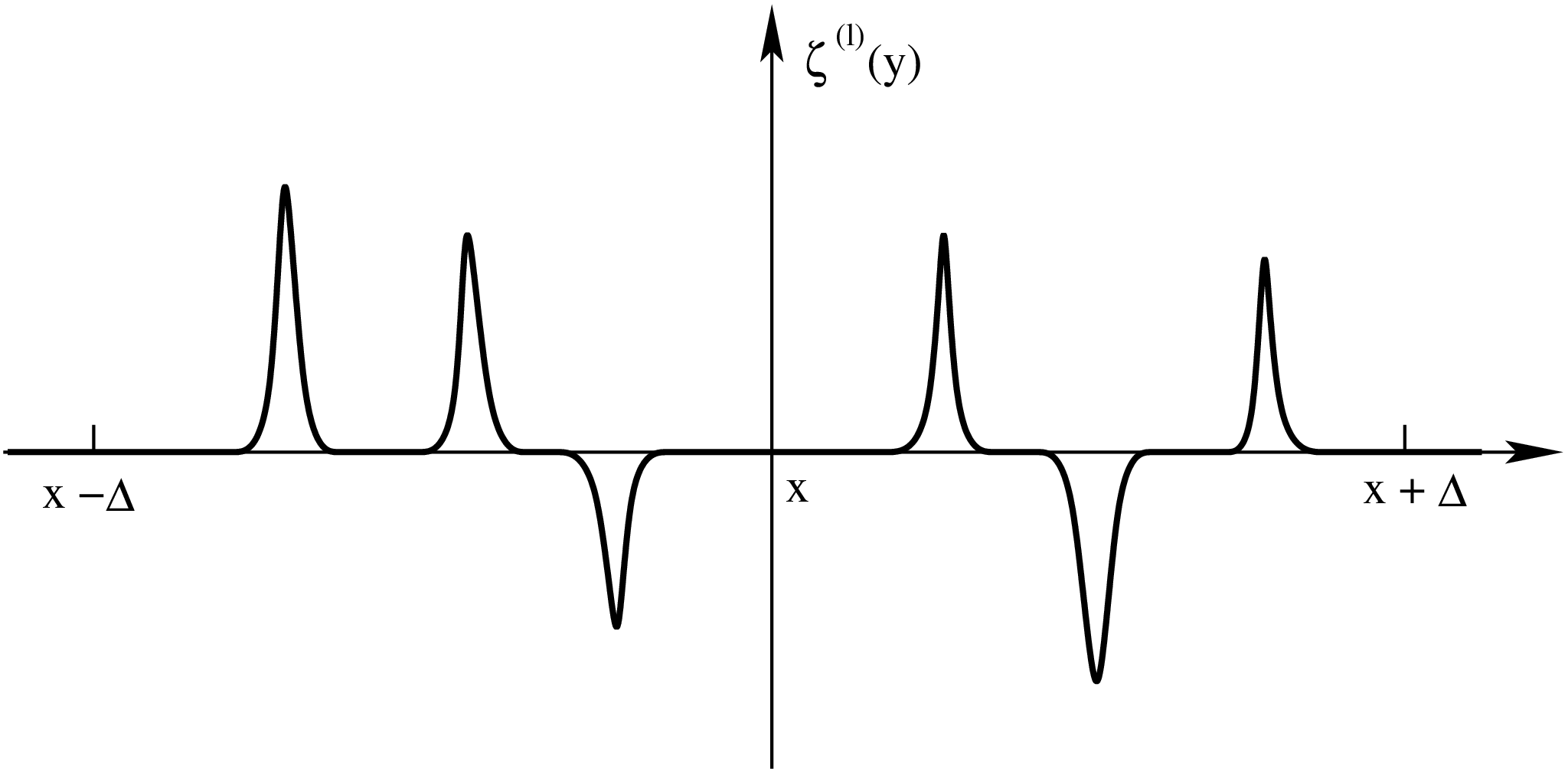}
\end{center}
\caption{The schematic possible form of the functions
$\, \zeta^{l} (y) \, $.}
\label{Functions}
\end{figure}

 Let us say again, that the functions 
$\, \zeta^{l} (y) \, $ can be easily constructed and it is 
most convenient to represent them in the form shown at
Fig. \ref{Functions}.

 Let us consider now the convolutions (in $y$) of the
full expression for
$\left[{\cal L}_{\xi} {\hat J}\right]^{ij}(x,y) \, $
with the infinitely smooth functions
$$\kappa_{j} (y) \,\, C^{l} \,\, 
{\hat \zeta}^{l} \left( x + C (y-x) \right) \quad ,
\quad \quad \quad l \,\, = \,\, 0 \, , \, \dots , 
\, Q - 1 $$
and put $\, C \, \rightarrow \, \infty \, $.

 Easy to see that the local part of
$\left[{\cal L}_{\xi} {\hat J}\right]^{ij}(x,y) \, $
will give us identical zero in such convolutions
due to the property (2) of the functions
$\, \zeta^{l} (y) \, $. In the same way, we will get
zero in the limit $\, C \, \rightarrow \, \infty \, $
in the non-local part (\ref{FirstTerm}) of
$\left[{\cal L}_{\xi} {\hat J}\right]^{ij}(x,y) \, $
according to the property (4) of the functions
$\, {\hat \zeta}^{l} (y) \, $. At the same time,
the non-local part (\ref{SecondTerm}) will give us
the values
$$\sum_{k\geq 0} \, e_{k} \, 
\left[ {1 \over 4} \int_{-\infty}^{\infty}
S^{p}_{(k)}(\varphi,\varphi_{w},\dots) \,
q_{p}(w) \, dw \right] \,\,\,
{1 \over 4} \, \sum_{s=1}^{Q} \, e^{\prime}_{s} \, 
A^{i}_{(s)} (\varphi,\varphi_{x},\dots) \,\,
a_{(s) lx} (x)  \,\,\, ,  
\quad \quad l \, = \, 0 \, , \, \dots , \, 
Q - 1 \,\, ,$$
in the limit $\, C \, \rightarrow \, \infty \, $
according to the property (3) of the functions
$\, \zeta^{l} (y) \, $.

 Coming back now to the property
$\,\, W (x) \neq 0 \,\, $ and assuming that in general
$$\sum_{k\geq 0} \, e_{k} \, 
\left[ {1 \over 2} \int_{-\infty}^{\infty}
S^{p}_{(k)}(\varphi,\varphi_{w},\dots) \,
q_{p}(w) \, dw \right] \,\,\, \neq \,\,\, 0 $$
for the linearly independent set 
$\, \{ {\bf S}_{(k)} \} \, $, we can claim now
that the vanishing of the expression
$\left[{\cal L}_{\xi} {\hat J}\right]^{ij}(x,y) \, $
implies in fact the relations
$\,\, A^{i}_{(s)} (\varphi,\varphi_{x},\dots) 
\, = \, 0 \,\, $ for our chosen function
$\,\, \varphi (y) \, \in \, {\cal S} \, $.
Using now the properties of the set $\, {\cal S} \, $
and the translational invariance of our Hamiltonian
operator we conclude now that
$\,\, A^{i}_{(s)} (\varphi,\varphi_{x},\dots) 
\, \equiv \, 0 \,\, $ on the full set of functions
which we consider. As a result, we can claim now that
the non-local part (\ref{SecondTerm}) of the expression
$\left[{\cal L}_{\xi} {\hat J}\right]^{ij}(x,y) \, $
is in fact identically equal to zero. As a consequence,
we can claim also the the non-local part 
(\ref{FirstTerm}) should be also identical zero on the
full set of functions $\, \varphi (x) \, $.

 Looking now at the form of the term (\ref{FirstTerm})
we can see that it represents a sum of linearly
independent tensor functions of $\, (x, y) \, $,
so we get that every coefficient, given by the integral
$${1 \over 2} \int_{-\infty}^{\infty} \, 
q_{p}(z) \left[ S_{(k)} ,S_{(k^{\prime})}\right]^{p}(z) 
\,\, dz \quad , $$
should be in fact identically equal to zero.
In view of the arbitrariness of the functions
$\, q_{p}(z) \, $ we obtain then
$$ \left[ S_{(k)} , S_{(k^{\prime})}\right]
\, \equiv \, 0 $$

  From (\ref{nulder1}) we then have also for a linearly
independent set of $S_{(k)}$ and different $q_{p}(w)$
that
 
$$ \left[{\cal L}_{k} {\hat J}\right]^{ij}(x,y) 
\, \equiv \, 0 $$

 So we obtain the statements of the theorem.

{\hfill {\it Theorem 1.1 is proved.}}

\vspace{5mm}

  It is also obvious that the statements of the theorem are
valid for all the brackets (\ref{nonlocbr}) written in the
``irreducible'' form, since all ${\tilde S}_{(k)}$ 
and ${\tilde T}_{(k)}$ in this case are just
linear combinations of the flows $S_{(k)}$ .

\vspace{1cm}

{\it Remark.}

 Let us point here that the first statement of the Theorem
for the non-local brackets (\ref{ferbr}) of
Ferapontov type was proved previously by E.V. Ferapontov 
in \cite{fer1} using differential-geometrical 
considerations. In \cite{fer1}-\cite{fer4} also the full
classification of the brackets (\ref{ferbr}) from the
differential geometrical point of view can be found.
 
 It is easy to see now that the local functional of type 
(\ref{hamilt}) 
$$I \, = \, \int {\cal P} (\varphi,\varphi_{x},\dots) \, dx $$
generates a local flow in the Hamiltonian structure 
(\ref{nonlocbr}) if and only if the derivative of its
density ${\cal P} (\varphi,\varphi_{x},\dots)$ with respect
to any of the flows (\ref{sflows}) represents total 
derivative with respect to $x$, i.e. there exist
such ${\cal Q}_{(k)} (\varphi,\varphi_{x},\dots)$ that
$${\cal P}_{\tau_{k}}(\varphi,\varphi_{x},\dots) \, \equiv \,
\partial_{x} {\cal Q}_{(k)} (\varphi,\varphi_{x},\dots)$$

 As was also pointed out by E.V.Ferapontov (\cite{fer1}), 
this means that the integral $I$ represents a conservation 
law for any of the systems (\ref{sflows}). 

 From the Theorem 1.1 we obtain now that the flows 
(\ref{sflows}) commute in fact with all the local Hamiltonian 
fluxes, generated by local functionals (\ref{hamilt}), since 
they conserve in this case both the Hamiltonian structure and 
the corresponding Hamiltonian functions.

\vspace{1cm}

\section{\bf The Whitham method and the ``regularity'' conditions.}
\setcounter{equation}{0}

 Now we come to Whitham's averaging procedure 
(see \cite{whith}-\cite{dm}).
Let us remind that in the $m$-phase Whitham's method for
systems (\ref{locsys}) we make a rescaling transformation
$X = \epsilon x$, $T = \epsilon t$ to obtain the system
\begin{equation}
\label{wsyst}
\epsilon \varphi^{i}_{T} \, = \, 
Q^{i}(\varphi, \epsilon \varphi_{X}, 
\epsilon^{2} \varphi_{XX},\dots)
\end{equation}

 Then we try to find functions
$$ S(X,T) \, = \, (S^{1}(X,T),\dots, S^{m}(X,T)) $$
and $2\pi$-periodic with respect to each $\theta^{\alpha} \, $
($\theta = (\theta^{1},\dots,\theta^{m})$) functions

$$\Phi^{i}(\theta, X,T,\epsilon) \, = \, \sum_{k=0}^{\infty}
\, \epsilon^{k} \, \Phi^{i}_{(k)}(\theta, X,T) \quad , $$
such that the functions 
\begin{equation}
\label{wsol}
\varphi^{i}(\theta, X,T,\epsilon) \, = \, \sum_{k=0}^{\infty}
\, \epsilon^{k} \,  
\Phi^{i}_{(k)} \Big(\theta + {S(X,T) \over \epsilon}, \, 
X, T \Big)
\end{equation}
satisfy system (\ref{wsyst}) at any $\theta$ in any order 
of $\epsilon$.

It follows then that $\Phi^{i}_{(0)}(\theta, X,T)$ at any $X$ 
and $T$ defines an exact $m$-phase solution of (\ref{locsys}),
depending on some parameters $U = (U^{1},\dots, U^{N})$ and
initial phases 
$\theta_{0} = (\theta^{1}_{0},\dots, \theta^{m}_{0})$ and,
besides that, we have the relations
$$S^{\alpha}_{T} = \omega^{\alpha}(U(X,T)) \,\,\, ,
\,\,\,\,\, S^{\alpha}_{X} = k^{\alpha}(U(X,T)) $$
where $\omega^{\alpha}(U)$ and $k^{\alpha}(U)$ are respectively 
the frequencies and the wave numbers of the corresponding
$m$-phase solution of (\ref{locsys}).

  The compatibility conditions of system (\ref{wsyst})
in the first order of $\epsilon$ together with the relations
$$ k^{\alpha}_{T} \, = \, \omega^{\alpha}_{X} $$
give us Whitham's system of equations on the parameters 
$U(X,T)$, which represents a quasi-linear system of hydrodynamic 
type

\begin{equation}
\label{whitheq}
U^{\nu}_{T} \, = \, V^{\nu}_{\mu}(U) \,\, U^{\mu}_{X}
\end{equation}

\vspace{5mm}

 Let us note here, that the representation of the modulated
solutions of system (\ref{locsys}) in the form (\ref{wsol})
is in fact usually possible just in the one-phase situation. 
In the multi-phase case we can usually write down just the 
first term in  the expansion (\ref{wsol}), while the higher 
order corrections have in general more complicated form
(see e.g. \cite{dobr1,dobr2,DobrKrichever}). Let us say,
however, that the Whitham system, defined as above, still
plays the central role in description of the modulated
solutions both in the one-phase and the multi-phase case.

\vspace{5mm}

  The first procedure of averaging of local 
field-theoretical Poisson brackets was proposed in
\cite{dn1}-\cite{dn3} by B.A. Dubrovin and S.P. Novikov.
This procedure permits to obtain local Poisson brackets
of Hydrodynamic type:
\begin{equation}
\label{locpbr}
\{U^{\nu}(X), U^{\mu}(Y)\} \,\,\, = \,\,\, g^{\nu\mu}(U) \,
\delta^{\prime}\, (X-Y) \,\, + \,\, 
b^{\nu\mu}_{\gamma}(U) \, U^{\gamma}_{X} \,\,
\delta (X-Y)
\end{equation}
for Whitham's system (\ref{whitheq}) from a local
Hamiltonian structure
$$\{\varphi^{i}(x), \varphi^{j}(y)\} \,\,\, = \,\,\, 
\sum_{k\geq 0} \, B^{ij}_{k}(\varphi,\varphi_{x},\dots)
\,\, \delta^{(k)}(x-y) $$
for the initial system (\ref{locsys}).

 The method of Dubrovin and Novikov is based on the presence
of $N$ (equal to the number of parameters $U^{\nu}$ of the 
family of $m$-phase solutions of (\ref{locsys})) local 
integrals
\begin{equation}
\label{integ}
I^{\nu} \,= \, 
\int {\cal P}^{\nu}(\varphi,\varphi_{x},\dots) \, dx
\quad ,
\end{equation}
commuting with the Hamiltonian function (\ref{hamilt}) and
with each other

\begin{equation}
\label{invv}
\{I^{\nu} , H\} \, = \, 0  \,\,\,\,\,  ,  \quad \quad 
\{I^{\nu} , I^{\mu}\} \, = \, 0  \,\,\,\,\,  , 
\end{equation}
and can be described in the following way:

\vspace{3mm}

  We calculate the pairwise Poisson brackets of the 
densities ${\cal P}^{\nu}$ in the form
$$\{{\cal P}^{\nu}(x), {\cal P}^{\mu}(y)\} \,\,\, = \,\,\,
\sum_{k\geq 0} \, A^{\nu\mu}_{k}(\varphi,\varphi_{x},\dots)
\,\, \delta^{(k)}(x-y) $$
where 
$$A^{\nu\mu}_{0}(\varphi,\varphi_{x},\dots) \, \equiv \,
\partial_{x} Q^{\nu\mu}(\varphi,\varphi_{x},\dots) $$
according to (\ref{invv}). Then the Dubrovin-Novikov bracket
on the space of functions $U(X)$ can be written in the form
\begin{equation}
\label{dubrnovb}
\{U^{\nu}(X), U^{\mu}(Y)\} \,\,\, = \,\,\,
\langle A^{\nu\mu}_{1}\rangle (U) \,\, \delta^{\prime}(X-Y) 
\, + \, {\partial \langle Q^{\nu\mu} \rangle \over 
\partial U^{\gamma}} \, U^{\gamma}_{X} \,\, \delta (X-Y) 
\end{equation}
where $\, \langle \dots \rangle$ means the averaging on the 
family of $m$-phase solutions of (\ref{locsys}) given by the 
formula: 
\footnote{Strictly speaking this formula is valid for 
generic set of the wave numbers $k^{\alpha}$, but 
we should use in any case the second
part of it  
for the averaged quantities to obtain the right procedure.
(Here $k^{\alpha}$ are continuous parameters on the family of 
the $m$-phase solutions).}
\begin{equation}
\label{usredfor}
\langle F \rangle \,\,\, = \,\,\, \lim_{c \rightarrow \infty}
\, {1 \over 2c} \, 
\int_{-c}^{c} F (\varphi, \, \varphi_{x}, \, \dots) \, 
dx  \,\,\, = \,\,\,
{1 \over (2\pi)^{m}} \int_{0}^{2\pi}\!\!\dots\int_{0}^{2\pi}
F \left(\Phi, \, k^{\alpha}(U) \, \Phi_{\theta^{\alpha}}, \,
\dots \right) \, d^{m}\theta
\end{equation}

 Here we choose the parameters $U^{\nu}$ such that they coincide
with the values of $I^{\nu}$ on the corresponding solutions
$$U^{\nu} \, = \, \langle P^{\nu}(x) \rangle $$

 The Jacobi identity for the averaged bracket (\ref{dubrnovb})
in the general 
case was proved in \cite{engam} (for systems having also
local Lagrangian formalism there was a proof in 
\cite{malpav}).

  Let us note here also that the procedure described above
gives a Poisson bracket only if we average the initial 
Hamiltonian structure on a ``full regular family'' of $m$-phase 
solutions (see \cite{novmal,engam,Sigma}).
We will formulate actually more precise requirements 
when describe the averaging procedure in the non-local case.

  Brackets (\ref{locpbr}) can be described from the 
differential-geometrical point of view. Thus, for a
non-degenerated tensor $\, g^{\nu\mu}$ we have in fact that 
it should represent a flat contravariant metric and the values
$$\Gamma^{\nu}_{\mu\gamma} \, = \, - g_{\mu\lambda} \,\,
b^{\lambda\nu}_{\gamma}$$
should give the Levi-Civita connection for the metric
$g_{\nu\mu}$ (with lower indices). The  brackets 
(\ref{locpbr}) with a degenerated tensor $\, g^{\nu\mu}$ 
are more complicated but also have a nice geometrical 
structure (see \cite{grinberg}). 

  The non-local Poisson brackets (\ref{ferbr}) give a  
generalization of local Poisson brackets of Dubrovin and 
Novikov and are closely connected with the integrability
of systems of hydrodynamic type, reducible to the diagonal 
form (\cite{tsarev}). Namely, any system reducible to the
diagonal form and Hamiltonian with respect to the bracket
(\ref{ferbr}) satisfies in fact (see \cite{fer1}-\cite{fer4})
the so-called ``semi-Hamiltonian'' property, introduced by
S.P. Tsarev (\cite{tsarev}), and can be integrated by
Tsarev's ``generalized hodograph method''. In \cite{bfer}
the investigation of possible equivalence of the
``semi-Hamiltonian'' properties introduced by Tsarev and
the Hamiltonian properties with respect to the bracket
(\ref{ferbr}) can be also found. 

 Let us also point out here that
the questions of integrability of Hamiltonian systems, 
which can not be written in the diagonal form, were studied 
in \cite{fer6}-\cite{fer9}.

\vspace{3mm}

  The procedure of averaging of the non-local Poisson brackets
in the Whitham method and the proof of the Jacobi identity for
the averaged non-local bracket resemble the same things 
for the local brackets. However the formulas of averaging 
and the proof contain in fact some essential differences,
so, we have to represent here special consideration 
for the non-local case. 

  The $m$-phase solutions of (\ref{locsys}) 
$$\varphi^{i}(x,t) \,\,\, = \,\,\, 
\Phi^{i}(\omega t + kx + \theta_{0}) \,\,\, , $$
where $\,\,\, \omega = (\omega^{1},\dots,\omega^{m})$,
$\,\, k = (k^{1},\dots,k^{m}) \, $, are defined by 
$2\pi$-periodic solutions of the system
\begin{equation}
\label{phasesys}
\omega^{\alpha} \, \Phi^{i}_{\theta^{\alpha}} \,\, - \,\,
Q^{i} \left(\Phi, \, k^{\alpha}\Phi_{\theta^{\alpha}}, \,
k^{\alpha}k^{\beta}\Phi_{\theta^{\alpha}\theta^{\beta}},
\, \dots \right) \,\,\, = \,\,\, 0 \,\,\, , 
\end{equation}
depending on $\omega$ and $k$ as on parameters. 
So, we assume that for generic $\, \omega$ and $\, k$
we obtain from (\ref{phasesys}) a finite-dimensional
submanifold $\, {\cal M}_{\omega,k}$ (in the space of 
$2\pi$-periodic with respect to each $\theta^{\alpha}$ 
functions), parameterized by the initial phase shifts 
$\, \theta^{\alpha}_{0}$ and maybe
also by some additional parameters $\, r^{1},\dots,r^{h}$.
\footnote{For the multiphase case ($m \geq 2$) it is
essential that the closure of any orbit generated by
the vectors $(\omega_{1},\dots,\omega_{m})$ and
$(k_{1},\dots,k_{m})$ in the $\theta$-space is the full
$m$-dimensional torus ${\bf T}^{m}$. For the case of
``rationally-dependent'' $\omega_{1},\dots,\omega_{m}$
and $k_{1},\dots,k_{m}$ and $m \geq 3$ we have that the operators
(\ref{phasesys}) are independent on each of such closed
submanifolds in ${\bf T}^{m}$ which can have the dimensionality
$ < m$. The functions from ${\cal M}_{\omega,k}$ can be found
in this case from the additional requirement that they define 
also $m$-phase solutions for systems (\ref{sflows}) 
(with some $\Omega_{(k)}^{\alpha}(\omega,k,r)$) and the systems
generated by the functionals $I^{\nu}$ (see later) (also with
some $\omega^{\alpha\nu}(\omega,k,r)$). All these requirements
uniquely define the finite-dimensional spaces
${\cal M}_{\omega,k}$, which continuously depend on the
parameters $\omega$ and $k$.}

 Combining all such ${\cal M}_{\omega,k}$ at different 
$\omega$ and $k$ we obtain that the $m$-phase solutions of the 
system  (\ref{locsys}) can be parameterized by $N = 2m + h$ 
parameters $\, U^{1},\dots,U^{N}$, invariant with respect to the 
initial shifts of 
$\, \theta^{\alpha}$, and the initial phase shifts 
$\, \theta^{\alpha}_{0}$ after the choice of some ``initial''
functions $\, \Phi^{i}_{(in)}(\theta, U)$, corresponding to the
zero initial phases. The joint of the submanifolds
${\cal M}_{\omega,k}$ at all $\omega$ and $k$ gives us a 
submanifold ${\cal M}$ in the space of $2\pi$-periodic with
respect to each $\, \theta^{\alpha}$ functions, which 
corresponds to the full family of $m$-phase solutions 
of (\ref{locsys}).

  For the Whitham procedure we should now require some 
``regularity'' properties of the system of constraints 
(\ref{phasesys}). Namely

\vspace{0.5cm}

{\it
  (I) We require that the linearized system (\ref{phasesys})
$${1 \over (2\pi)^{m}} \int_{0}^{2\pi}\!\!\dots\int_{0}^{2\pi}
\left( \, \omega^{\alpha} \,\, \delta^{i}_{j}  \,\, 
\delta_{\theta^{\alpha}} (\theta - \theta^{\prime}) \,\,\, 
- \,\,\, {\delta Q^{i} (\theta) \over \delta \varphi^{j} 
(\theta^{\prime})} \, \right) \,\, \Psi^{j} (\theta^{\prime})
\,\, d^{m} \theta^{\prime} \,\,\, = \,\,\, 0 $$
has for generic $\omega$ and $k$ exactly 
$\, h+m = N-m$ solutions (``right eigen vectors'') 
$\, \xi_{(q) \omega,k}(\theta, r)$
at the corresponding ``points'' of ${\cal M}_{\omega,k}$,
given by the vectors tangential
to $\, {\cal M}_{\omega,k}$, i.e. the functions
$\, \Phi_{\theta^{\alpha}}(\theta,r,\omega,k)$ and
$\, \Phi_{r^{q}}(\theta,r,\omega,k)$
(at the fixed values of $\omega$ and $k$). }

\vspace{0.5cm}

{\it
  (II) We also require that the number of linearly
independent ``left eigen vectors'' 
$\, \kappa_{(q) \omega,k}(\theta, r)$, orthogonal to the
image of the introduced linear operator,  
is exactly the same
($N-m$) as the number of the ``right eigen vectors''
$\, \xi_{(q) \omega,k}(\theta, r)$ for generic 
$\omega$ and $k$. In addition, we will assume that
the corresponding
$\, \kappa_{(q) \omega,k}(\theta, r)$ also depend
continuously on the parameters $U^{\nu}$ on ${\cal M}$. }

\vspace{0.5cm}

  The requirements (I) and (II) are actually closely
connected with the Whitham procedure and the asymptotic
solutions (\ref{wsol}). Indeed, it is not difficult
to see that every $k$-th term in the expansion (\ref{wsol})
is determined by the defined above linear system with a
nontrivial right-hand part, depending on the previous terms 
of (\ref{wsol}). For resolvability of these systems we
have in any case to require the orthogonality of the 
right-hand part to all the ``regular left eigen vectors''
$\, \kappa_{(q) \omega,k}(\theta, r)$, corresponding
to the zero eigen values. The corresponding orthogonality 
conditions in the first order of $\epsilon$ together with 
the relations
$$k_{T} \, = \, \omega_{X}$$
give a system of $\,\, (N - m) + m \, = \, N \, $
equations, which coincides (by definition) with the 
Whitham's system of equations (\ref{whitheq}).

  Let us now discuss the requirements (I) and (II)
from the Hamiltonian point of view.

\vspace{0.5cm}

  First of all, like in the procedure of Dubrovin
and Novikov, for the procedure of averaging of the bracket
(\ref{canform}) we need a set of integrals
$\, I^{\nu} , \,\, \nu = 1,\dots,N $, satisfying the 
following requirements:

\vspace{3mm}

  (A) Every $I^{\nu}$ is a local functional
\begin{equation}
\label{inu}
I^{\nu} \,\,\, = \,\,\, \int 
{\cal P}^{\nu}(\varphi,\varphi_{x},\dots) \, dx \,\,\, ,
\end{equation}
which generates a local flow
\begin{equation}
\label{nuflows}
\varphi^{i}_{t^{\nu}} \,\,\, = \,\,\,
Q^{i}_{(\nu)}(\varphi,\varphi_{x},\dots)
\end{equation}
with respect to the bracket (\ref{canform}).

\vspace{3mm}

 As was pointed above we should require then that the local
flows (\ref{sflows}), defined by the bracket (\ref{canform})
in the ``canonical'' (or ``irreducible'') form, conserve all 
the $I^{\nu}$, i.e. the time derivatives of the corresponding 
${\cal P}^{\nu}(\varphi,\varphi_{x},\dots)$ with respect to each
of the flows (\ref{sflows}) represent total derivatives with 
respect to $x$

\begin{equation}
\label{pfder}
{d \over dt^{k}} \, {\cal P}^{\nu}(\varphi,\varphi_{x},\dots)
\,\,\, \equiv \,\,\,  
\partial_{x} F^{\nu}_{(k)}(\varphi,\varphi_{x},\dots)
\end{equation}
for some functions $F^{\nu}_{(k)}(\varphi,\varphi_{x},\dots)$.

\vspace{3mm}

 (B) All $I^{\nu}$ commute with each other and with the 
Hamiltonian function (\ref{hamilt})

\begin{equation}
\label{coms}
\{I^{\nu},I^{\mu}\} = 0 \quad , \quad \quad \{I^{\nu},H\} = 0
\end{equation}

\vspace{3mm}

 (C) The averaged densities $\langle{\cal P}^{\nu}\rangle$
\begin{equation}
\label{avden}
\langle{\cal P}^{\nu}\rangle \,\,\, = \,\,\, 
\lim_{c \rightarrow \infty} \, {1 \over 2c} \, \int_{-c}^{c} 
{\cal P}^{\nu} (\varphi, \, \varphi_{x}, \, \dots) 
\, dx \,\,\, = \,\,\,
{1 \over (2\pi)^{m}}\int_{0}^{2\pi}\!\!\dots\int_{0}^{2\pi}
{\cal P}^{\nu} \left(\Phi, \, k^{\alpha}\Phi_{\theta^{\alpha}},
\, \dots \right) \, d^{m}\theta
\end{equation}
can be regarded as independent coordinates 
$\, U^{1},\dots,U^{N}$ on the family of $m$-phase solutions of
(\ref{locsys}).\footnote{Here again we can use everywhere
the second part of the formula (\ref{avden}) for the averaged 
values on ${\cal M}$.}

\vspace{3mm}

 From the requirements above we immediately obtain that the
flows (\ref{nuflows}) commute with our initial system 
(\ref{locsys}) and with each other.

 From Theorem 1.1 we obtain also that the commutative flows
(\ref{sflows}), defined by the Poisson bracket, also commute with
(\ref{locsys}) and (\ref{nuflows}) since they conserve the 
corresponding Hamiltonian functions and the Hamiltonian structure
(\ref{canform}).

 Now we can consider the functionals
\begin{equation}
\label{perfunnu}
{\bar I}^{\nu} \,\,\, = \,\,\, 
\lim_{c \rightarrow \infty} \, {1 \over 2c} \,
\int_{-c}^{c} {\cal P}^{\nu} (\varphi,\varphi_{x},\dots) \, dx
\end{equation}
and
\begin{equation}
\label{perham}
{\bar H} \,\,\, = \,\,\, 
\lim_{c \rightarrow \infty} \, {1 \over 2c} \,
\int_{-c}^{c} {\cal P}_{H} (\varphi,\varphi_{x},\dots) \, dx
\end{equation}
on the space of the quasiperiodic functions (with $m$
wave numbers).

\vspace{0.5cm}

{\it It is easy to see now 
that the local fluxes (\ref{locsys}),
(\ref{sflows}) and (\ref{nuflows}), being considered on the
space of the quasiperiodic functions, also conserve the values
of ${\bar I}^{\nu}$ and ${\bar H}$ and commute with each other,
since these properties can be expressed just as local
relations containing $\, \varphi,\varphi_{x},\dots$ and the
time derivatives of the densities
$\, {\cal P}^{\nu} (\varphi,\varphi_{x},\dots)$,
$\, {\cal P}_{H} (\varphi,\varphi_{x},\dots)$ 
at the same point $x$.}

\vspace{0.5cm}

 Now we can conclude that all the fluxes (\ref{sflows})
and (\ref{nuflows}) leave invariant the family of $m$-phase
solutions, given by (\ref{phasesys}), and can generate on it 
only linear shifts of the initial phases 
$\, \theta^{\alpha}_{0}$, which follows from the commutativity 
of the flows
\begin{equation}
\label{sperfl}
\varphi^{i}_{\tau^{k}} (\theta) \,\,\, = \,\,\, S^{i}_{(k)}
\left(\varphi, \, k^{\alpha}\varphi_{\theta^{\alpha}}, \,
k^{\alpha}k^{\beta}\varphi_{\theta^{\alpha}\theta^{\beta}},
\, \dots \right)
\end{equation}
and
\begin{equation}
\label{nuperfl}
\varphi^{i}_{t^{\nu}} (\theta) \,\,\, = \,\,\, Q^{i}_{(\nu)}
\left(\varphi, \, k^{\alpha}\varphi_{\theta^{\alpha}}, \,
k^{\alpha}k^{\beta}\varphi_{\theta^{\alpha}\theta^{\beta}},
\, \dots \right)
\end{equation}
with the flows 
$\, \varphi^{i}_{t^{\alpha}} = \varphi^{i}_{\theta^{\alpha}}$
and
$$\varphi^{i}_{t} \,\,\, = \,\,\, Q^{i}
\left( \varphi, \, k^{\alpha}\varphi_{\theta^{\alpha}}, \,
k^{\alpha}k^{\beta}\varphi_{\theta^{\alpha}\theta^{\beta}},
\, \dots \right)$$
on the space of $2\pi$-periodic with respect to each
$\theta^{\alpha}$ functions and the conservation of the 
functionals ${\bar I}^{\nu}$ (i.e. $U^{\nu}$ on ${\cal M}$)
by the flows (\ref{sperfl}) and (\ref{nuperfl}). 
(Here $k^{\alpha}$ are $m$ wave numbers
of the function $\varphi(x)$.) So, we obtain
that our family of $m$-phase solutions of (\ref{locsys})
represents also a family of $m$-phase solutions for systems
(\ref{sflows}) and (\ref{nuflows}), and we can consider
also the Whitham equations for these systems, based on the
family ${\cal M}$.

 We can also conclude that in our situation the variational 
derivatives of the functionals (\ref{perfunnu}) and 
(\ref{perham}) with respect to $\varphi(\theta)$ at the points
of the submanifold ${\cal M}$ represent some linear 
combinations of the corresponding ``left eigen vectors''
$\, \kappa_{(q)}(\theta + \theta_{0}, U)$ (see 
\cite{dn2}-\cite{dm} and references therein). 
Indeed, from the 
conservation of the functionals (\ref{perfunnu}) and
(\ref{perham}) by the flows 
$\, \varphi^{i}_{t^{\alpha}} = \varphi^{i}_{\theta^{\alpha}}$
and
$$\varphi^{i}_{t} \,\,\, = \,\,\, Q^{i} \left(\varphi, \, 
k^{\alpha}\varphi_{\theta^{\alpha}}, \, \dots \right)$$
we can conclude that the convolution of their derivatives
(with respect to $\varphi^{i}(\theta)$) with the system of
constraints (\ref{phasesys}) is identically zero for all the
periodic functions with respect to all $\theta^{\alpha}$
and for any $k^{1},\dots,k^{m}$ and
$\omega^{1},\dots,\omega^{m}$. So we can take the variational
derivative of the corresponding expression with respect
to $\varphi^{j}(\theta^{\prime})$ and then omit the second 
variational derivative of $\, {\bar I}^{\nu}$ and 
$\, {\bar H}$ according to the conditions (\ref{phasesys}).
After that we obtain that the variational derivatives
of $\, {\bar I}^{\nu}$ and ${\bar H}$
are also orthogonal to the image of the linearized operator
(\ref{phasesys}) at the points of ${\cal M}$
and so represent some linear combinations
of $\, \kappa_{(q)}(\theta + \theta_{0}, U)$ on
${\cal M}$.

\vspace{0.5cm}

{\bf Lemma 2.1}

{\it
 Suppose we have the properties (A)-(C) and (I)-(II) for
our family of $m$-phase solutions of (\ref{locsys}). Let us
put
\begin{equation}
\label{upi}
U^{\nu} \,\,\, = \,\,\, \langle{\cal P}^{\nu}\rangle 
\,\,\, = \,\,\, {\bar I}^{\nu}
\end{equation}
on the space ${\cal M}$ and define the functions
$\, k^{\alpha} = k^{\alpha}(U)$ on the submanifold ${\cal M}$.

 Then the functionals 
$\, K^{\alpha} = k^{\alpha}({\bar I}[\varphi])$ on the space of
$2\pi$-periodic with respect to each $\theta^{\alpha}$ functions
(and also on the space of quasiperiodic functions $\varphi(x)$
with $m$ wave numbers) have zero variational derivatives on
the submanifold ${\cal M}$. }

\vspace{0.5cm}

{\it Proof.}

 As we have from (II), the maximal number of linearly
independent variational derivatives of $\, {\bar I}^{\nu}$ on
${\cal M}$ is equal to $\, h+m = N-m$. So, we have $m$ linearly 
independent relations
\begin{equation}
\label{intder}
\sum_{\nu=1}^{N} \, \lambda^{\alpha}_{\nu}(U) \, 
{\delta {\bar I}^{\nu} \over \delta \varphi(\theta)} 
\,\,\, \equiv \,\,\, 0 \quad , \quad \quad
\alpha = 1,\dots,m
\end{equation}
($\varphi(\theta) = 
(\varphi^{1}(\theta),\dots,\varphi^{m}(\theta))$),
considered at given $\, k^{1},\dots,k^{m}$ at any point
of ${\cal M}$
(or in other words
\begin{equation}
\label{intder1}
\sum_{\nu=1}^{N} \, \lambda^{\alpha}_{\nu}(U) \,
{\delta {\bar I}^{\nu} \over \delta \varphi(x)} 
\,\,\, \equiv \,\,\, 0  \quad ,  \quad \quad
\alpha = 1,\dots,m
\end{equation} 
when considered on the space of functions with $m$ wave 
numbers.) We can use here the standard expression for the 
variational derivative and the formula (\ref{avden}) for 
$\, {\bar I}^{\nu}$. 

 Since we can obtain a change of the values of these linear 
combinations of the functionals $\, {\bar I}^{\nu}$ on 
$\, {\cal M}$ only due to variations of $k$ in (\ref{avden}) 
but not of $\, \varphi^{i}(\theta)$ (or in other words only if 
we have non-bounded variations of $\, \varphi^{i}(x)$ after
the variations of the wave numbers) we have on ${\cal M}$ 
\begin{equation}
\label{lammudk}
\sum_{\nu=1}^{N} \, \lambda^{\alpha}_{\nu}(U) \, dU^{\nu} 
\,\,\, = \,\,\,
\sum_{\beta=1}^{m} \, \mu^{(\alpha)}_{\beta}(U) \, 
dk^{\beta}(U)
\end{equation}
for some functions $\, \mu^{(\alpha)}_{\beta}(U)$.

 Since $\, U^{\nu}$ represent independent coordinates on 
$\, {\cal M}$, the matrix $\, \mu^{(\alpha)}_{\beta}$ has 
the full rank and is reversible. So, we get the
differentials $\, dk^{\beta}$ as some linear combinations of
differentials 
$\, \sum_{\nu=1}^{N} 
\lambda^{\alpha}_{\nu}(U) \, dU^{\nu} $,
corresponding to the functionals with zero derivatives on
$\, {\cal M}$
$$dk^{\beta} \,\,\, = \,\,\, \sum_{\alpha=1}^{m} \,
(\mu^{-1})^{\beta}_{(\alpha)} \,\,
\sum_{\nu=1}^{N} \, \lambda^{(\alpha)}_{\nu}(U) \,
dU^{\nu}$$ 

 So Lemma 2.1 now follows from (\ref{intder}).

\vspace{0.5cm}

{\it Remark 1.}

 As can be seen from the proof of Lemma 2.1, the variational
derivatives of $\, {\bar I}^{\nu}$ on ${\cal M}$ should span 
the full $\,(N-m)$-dimensional linear space, generated by all 
$\, \kappa_{(q)}(\theta + \theta_{0}, U)$, if we want to take
$\, \langle{\cal P}^{\nu}\rangle$ as a set of independent 
coordinates on ${\cal M}$. It is essential also that we consider 
the full family of $m$-phase solutions, given by (\ref{phasesys})
at different $\omega$ and $k$, (but not its ``submanifold'') 
and have $m$ independent relations (\ref{lammudk}) on $N$ 
differentials $\, dU^{\nu}$ from $m$ relations (\ref{intder}).

\vspace{0.5cm}

{\it Remark 2.}

 Let us note here that the equations (\ref{intder1}) were
introduced at first by S.P. Novikov in \cite{novikov}
as the definition of the $m$-phase solutions for the
KdV equation.

\vspace{0.5cm}

 Let us now prove a technical lemma which we will need later.

\vspace{0.5cm}

{\bf Lemma 2.2}

{\it
 Let us introduce the additional densities
\begin{equation}
\label{adden}
\Pi^{\nu}_{i(k)}(\varphi,\varphi_{x},\dots) 
\,\,\, \equiv \,\,\,
{\partial {\cal P^{\nu}}(\varphi,\varphi_{x},\dots) \over
\partial \varphi^{i}_{kx}}
\end{equation}
for $k \geq 0$, where 
$\, \varphi^{i}_{kx} \, \equiv \, 
\partial^{k} \varphi^{i} /\partial x^{k}$.

 Then on the submanifold ${\cal M}$ we have the relation
\begin{multline}
\label{kuprel}
\sum_{\nu=1}^{N} \, 
{\partial k^{\alpha} \over \partial U^{\nu}} \, 
{1 \over (2\pi)^{m}}\int_{0}^{2\pi}\!\!\dots\int_{0}^{2\pi}
\sum_{p\geq 1} \, p \,\, 
k^{\beta_{1}}(U)\dots k^{\beta_{p-1}}(U) \,\, 
\Phi^{i}_{(in)\theta^{\beta}\theta^{\beta_{1}}
\dots\theta^{\beta_{p-1}}}(\theta,U) \, \times   \\
\times \, \Pi^{\nu}_{i(p)} \left( \Phi_{(in)}(\theta,U),
\, k^{\gamma}\Phi_{(in)\theta^{\gamma}}(\theta,U), 
\, \dots \, \right) \,\, d^{m}\theta 
\quad \equiv \quad \delta^{\alpha}_{\beta}
\end{multline}
at any $U$ and $\theta_{0}$. }

\vspace{0.5cm}

{\it Proof.}

 According to Lemma 2.1 we should not take into account 
variations of the form of 
$\, \Phi_{(in)}(\theta + \theta_{0},U)$
when we consider infinitesimal changes of the values of the 
functionals $k^{\alpha}({\bar I})$ on ${\cal M}$. So, the only 
source for a change of these functionals on ${\cal M}$ is the 
dependence on the wave numbers $k$ in the expressions
$${\bar I}^{\nu} \,\,\, = \,\,\,
{1 \over (2\pi)^{m}}\int_{0}^{2\pi}\!\!\dots\int_{0}^{2\pi}
{\cal P}^{\nu} \left(
\Phi_{(in)}, \, k^{\gamma}\Phi_{(in)\theta^{\gamma}}, \,
k^{\gamma}k^{\delta}\Phi_{(in)\theta^{\gamma}\theta^{\delta}},
\, \dots \right) \, d^{m}\theta $$

 So, we can write
$$d\left(k^{\alpha}({\bar I})|_{{\cal M}}\right) \,\,\, = \,\,\,
\sum_{\nu=1}^{N} \,\, 
{\partial k^{\alpha} \over \partial U^{\nu}}(U) \,\,
{\partial {\bar I}^{\nu}[\varphi] \over \partial k^{\beta}}
|_{{\cal M}} \,\, dk^{\beta} $$
where the values of 
$\, \partial {\bar I}^{\nu}[\varphi]/\partial k^{\beta}$
on ${\cal M}$ are given by the integral expressions from 
(\ref{kuprel}). Since the values of the functionals
$\, k^{\alpha}({\bar I})$ on ${\cal M}$ coincide 
by definition with the wave numbers $k^{\alpha}$, we obtain 
the relation (\ref{kuprel}).

{\hfill {\it Lemma 2.2 is proved.}}

\vspace{0.5cm}
 
 For the evolution of the densities 
$\, {\cal P}^{\nu}(\varphi,\varphi_{x},\dots)$ according to 
our system (\ref{locsys}) we can also write the relations
\begin{equation}
\label{phlaw}
{d \over dt} \, {\cal P}^{\nu}(\varphi,\varphi_{x},\dots) 
\,\,\, \equiv \,\,\,
\partial_{x} Q^{\nu H}(\varphi,\varphi_{x},\dots)
\end{equation}
and the Whitham's system (\ref{whitheq}) can be also written
in the following ``conservative'' form

\begin{equation}
\label{conswhit}
\partial_{T} \, U^{\nu} \, = \,
\partial_{X} \langle Q^{\nu H}\rangle
\quad ,  \quad \quad 
\nu = 1,\dots,N
\end{equation}
for the parameters 
$\, U^{\nu} \, = \, \langle {\cal P}^{\nu} \rangle $, which 
gives an equivalent form of the Whitham equations. 

\vspace{3mm}

 The conservative form (\ref{conswhit}) of the Whitham's 
system will be very convenient in our considerations
of the averaging of Hamiltonian structures.
 
\vspace{3mm}
 
 Let us now put some ``regularity'' requirements about 
the joint ${\cal M}$ of the submanifolds ${\cal M}_{\omega,k}$ 
for all $\omega$ and $k$, corresponding to the full set of the
$m$-phase quasiperiodic solutions of the system (\ref{locsys}).

\vspace{0.5cm}

{\it
 (III) We require that ${\cal M}$ represents an 
$\, (N+m)$-dimensional submanifold in the space of the 
$2\pi$-periodic with respect to each 
$\, \theta^{\alpha}$ functions. }

\vspace{0.5cm}

 The property (III) means nothing but the fact that the shapes
of the solutions of (\ref{phasesys}) are all different at
different $\omega$ and $k$ in the space of the $2\pi$-periodic
vector-functions of $\theta$ so that $\omega$ and $k$ can be
reconstructed from them. It is easy to see that this 
requirement corresponds to the generic situation. We will
use here the property (III) in our procedure of averaging of
bracket (\ref{nonlocbr}).

\vspace{0.5cm}

 We will work with the full family of $2\pi$-periodic
solutions of (\ref{phasesys}) which will also depend 
on the ``slow'' variables $X$ and $T$. To define the
corresponding submanifold in the space of functions 
$\, \varphi (\theta,X,T)$ we should
extend the coordinates $U^{\nu}$ as functionals of
$\varphi (\theta)$ in the vicinity of our submanifold
${\cal M}$. This can be easily done (see \cite{engam}) in the
following way:

 Let introduce $N$ different functionals
$$A^{\nu} \,\,\, = \,\,\,
{1 \over (2\pi)^{m}}\int_{0}^{2\pi}\!\!\dots\int_{0}^{2\pi}
a^{\nu}(\varphi, \varphi_{\theta^{\alpha}},
\varphi_{\theta^{\alpha}\theta^{\beta}},\dots) 
\, d^{m}\theta  \quad , $$
such that their values ${\bar A}^{\nu}$ are functionally
independent on the functions from the submanifold ${\cal M}$.
Then we can express $U^{\nu} = U^{\nu}({\bar A})$ in terms of 
${\bar A}^{\nu}$ on ${\cal M}$ and after that extend them as the
functionals $U^{\nu}(A)$ on the space of $2\pi$-periodic with
respect to each $\theta^{\alpha}$ functions. 

  We can also expand
the coordinates $\theta^{\alpha}_{0}$ (see \cite{engam}) in the
vicinity of ${\cal M}$ by introduction of, say, functionals
$$B^{\alpha}[\varphi(\theta)] \,\,\, = \,\,\, 
{1 \over (2\pi)^{m}}\int_{0}^{2\pi}\!\!\dots\int_{0}^{2\pi}
\varphi_{\theta^{\alpha}}(\theta) \, \Phi_{(in)}
\left( \theta,U[\varphi] \right) \, d^{m}\theta  \quad , $$
which have zero values for 
$\varphi (\theta) = \Phi_{(in)}(\theta,U[\varphi])$.
In the generic situation we can locally express the values of
$\theta^{\alpha}_{0}$ on ${\cal M}$ in terms of ${\bar B}^{\alpha}$
and after that put
$\theta^{\alpha}_{0} = \theta^{\alpha}_{0}(B[\varphi])$ in the
corresponding local coordinate maps in the vicinity of
${\cal M}$.

  Now we consider the system
\begin{equation}
\label{fullsyst}
\varphi^{i}(\theta,X) \,\, - \,\, \Phi^{i}_{(in)}
\left( \theta + \theta_{0}[\varphi], \, U[\varphi] \right)
\,\,\, \equiv \,\,\, 0  \,\,\,  ,
\end{equation}
where $\theta^{\alpha}_{0}[\varphi]$ and
$U^{\nu}[\varphi]$ are the functionals in the vicinity
of ${\cal M}$, as a system of constraints, which defines
${\cal M}$ in the space of $2\pi$-periodic with respect to
each $\theta^{\alpha}$ functions.

\vspace{0.5cm}

 We can see now that the linearized system (\ref{fullsyst})
$${1 \over (2\pi)^{m}}\int_{0}^{2\pi}\!\!\dots\int_{0}^{2\pi}
\left(L^{i}_{j [U,\theta_{0}]}(\theta,\theta^{\prime}) \, 
\delta \Phi^{j}(\theta^{\prime})\right) d^{m}\theta^{\prime}  
\,\,\, = \,\,\, 0 \,\,\, , $$
where
\begin{multline*}
L^{i}_{j[U,\theta_{0}]}(\theta,\theta^{\prime}) 
\,\,\, \equiv \,\,\,
\delta^{i}_{j} \, 
\delta (\theta - \theta^{\prime}) \,\,\, -   \\
- \,\,\, \sum_{\alpha = 1}^{m}
\Phi^{i}_{(in)\theta^{\alpha}} \left( \theta + 
\theta_{0}[\varphi], U[\varphi] \right) \times 
{\delta \theta^{\alpha}_{0}[\varphi] \over \delta
\varphi^{j}(\theta^{\prime})} \,\,\, - \,\,\, 
\sum_{\nu = 1}^{N}
\Phi^{i}_{(in)U^{\nu}} \left( \theta +
\theta_{0}[\varphi], U[\varphi]\right) \times 
{\delta U^{\nu}_{0}[\varphi] \over \delta
\varphi^{j}(\theta^{\prime})}  \quad ,
\end{multline*}
has at any point $(U,\theta_{0})$ of ${\cal M}$ exactly $N+m$
solutions $\, {\tilde {\xi}}_{(p)[U,\theta_{0}]}(\theta)$,
corresponding to the tangential to ${\cal M}$ vectors
$\, \Phi_{(in)\theta^{\alpha}}$ and $\, \Phi_{(in)U^{\nu}}$, 
$\,\, \alpha = 1,\dots,m $,  $\, \nu = 1,\dots,N$. 

\vspace{0.5cm}

 It is evident also that all the ``left eigen vectors'' 
$\, {\tilde \kappa}_{(p)[U,\theta_{0}]}(\theta)$, orthogonal 
to the image of ${\hat L}$, are given by the variational 
derivatives 
$\, \delta \theta^{\alpha}_{0}[\varphi]/\delta
\varphi^{j}(\theta)$ and 
$\, \delta U^{\nu}_{0}[\varphi]/\delta
\varphi^{j}(\theta)$.

\vspace{0.5cm}

 From the invariance of the submanifold ${\cal M}$ with respect
to the flows (\ref{sflows}) and (\ref{nuflows}) we can also
write here the relations
\begin{equation}
\label{ssviaz}
{1 \over (2\pi)^{m}}\int_{0}^{2\pi}\!\!\dots\int_{0}^{2\pi}
L^{i}_{j [U,\theta_{0}]}(\theta, \, \theta^{\prime}) \,\,
S^{j}_{(k)} \left( \Phi_{(in)}(\theta^{\prime} + 
\theta_{0},U), \, k^{\alpha}
\Phi_{(in)\theta^{\alpha}}(\theta^{\prime} + 
\theta_{0},U), \dots \right) \, 
d^{m}\theta^{\prime} \,\,\, \equiv \,\,\, 0
\end{equation}
and
\begin{equation}
\label{nusviaz}
{1 \over (2\pi)^{m}}\int_{0}^{2\pi}\!\!\dots\int_{0}^{2\pi} 
L^{i}_{j [U,\theta_{0}]}(\theta,\theta^{\prime}) \,\,
Q^{j}_{(\nu)}(\Phi_{(in)}
\left( \theta^{\prime} + \theta_{0},U), \, k^{\alpha}
\Phi_{(in)\theta^{\alpha}}(\theta^{\prime} + \theta_{0},U),
\, \dots \right) \, d^{m}\theta^{\prime} \,\,\, \equiv \,\,\, 0
\end{equation}
for any $i, k$ and $\nu$ at any point $(U,\theta_{0})$ 
of ${\cal M}$, where $k^{\alpha} = k^{\alpha}[\Phi]$ can be 
considered now as the values of the corresponding functionals 
on ${\cal M}$.

 We now introduce the space of functions $\varphi(\theta,X,T)$,
depending on ``slow'' parameters $X$ and $T$ and $2\pi$-periodic 
with respect to each $\theta^{\alpha}$. Systems (\ref{fullsyst}),
considered independently at different $X$, give us a system 
of constraints defining the submanifold ${\cal M}^{\prime}$
in the space of functions $\varphi(\theta,X)$, corresponding to 
$m$-phase solutions of (\ref{locsys}) depending on the
additional parameters $X$ and $T$. 

 It will be actually convenient to introduce also the ``modified''
constraints (\ref{fullsyst})
\begin{equation}
\label{svgi}
G^{i}_{[U,\theta_{0}]}(\theta,X) \,\,\, = \,\,\, 
{1 \over (2\pi)^{m}}\int_{0}^{2\pi}\!\!\dots\int_{0}^{2\pi}
L^{i}_{j [U,\theta_{0}]}(\theta,\theta^{\prime}) \,
\left(\varphi^{j}(\theta^{\prime}) \, - \,  
\Phi^{j}_{(in)}(\theta^{\prime} + \theta_{0}[\varphi],U[\varphi])
\right) d^{m}\theta^{\prime}
\end{equation}
and take $\, U^{\nu}(X)$, $\theta^{\alpha}_{0}(X)$ and
$\, G^{i}_{[U,\theta_{0}]}(\theta,X)$, such that
\begin{equation}
\label{restr}
{1 \over (2\pi)^{m}}\int_{0}^{2\pi}\!\!\dots\int_{0}^{2\pi}
{\tilde \kappa}_{(p)[U]}(\theta + \theta_{0}(X)) \,\,
G^{i}_{[U,\theta_{0}]}(\theta,X) \,\, d^{m}\theta 
\,\,\, \equiv \,\,\, 0 
\quad ,  \quad \quad  p = 1, \dots, N+m  \quad ,
\end{equation}
as coordinates in the vicinity of ${\cal M}^{\prime}$
instead of the $\varphi^{i}(\theta,X)$. It is easy to see
also that we can find uniquely $\varphi^{i}(\theta,X)$ from
the relations 
$${1 \over (2\pi)^{m}}
\int_{0}^{2\pi}\!\!\!\!\!\dots\int_{0}^{2\pi}\!\!\!
L^{i}_{j [U,\theta_{0}]}(\theta,\theta^{\prime})
\left(\varphi^{j}(\theta^{\prime}) \, - \, 
\Phi^{j}_{(in)}(\theta^{\prime} + \theta_{0},U)
\right) d^{m}\theta^{\prime} \,\,\, = \,\,\, 
G^{i}_{[U,\theta_{0}]}(\theta,X)$$
and the values of
$\, U^{\nu}(X)$ and $\, \theta^{\alpha}_{0}(X)$
under the conditions (\ref{restr}).
\footnote{This system of constraints is different from
the system introduced in \cite{engam}.}
\vspace{0.5cm}

{\it Remark.}

Certainly we have here some freedom in the choice of the 
constraints $G^{i}(\theta,X)$. For example we can take also
the expressions (\ref{fullsyst}) 
as a  system of constraints 
defining ${\cal M}^{\prime}$. We 
prefer here to take the constraints in the form (\ref{svgi})
just to fix the uniform orthogonality conditions
(\ref{restr}) in the vicinity of ${\cal M}^{\prime}$.

\vspace{1cm}

 We will need also another coordinate system in the vicinity of
${\cal M}^{\prime}$, which differs from the described above by a
transformation, depending on the small parameter $\epsilon$
and singular at $\epsilon \rightarrow 0$. Namely, we recall our
integrals (\ref{integ})
$$I^{\nu} \,\,\, = \,\,\, 
\int {\cal P}^{\nu}(\varphi,\varphi_{x},\dots) \, dx 
\,\,\, , $$
make a transformation $\, X = \epsilon x$ and define the
functionals
\begin{equation}
\label{jint}
J^{\nu}(X) \,\,\, = \,\,\, 
{1 \over (2\pi)^{m}}\int_{0}^{2\pi}\!\!\dots\int_{0}^{2\pi}
{\cal P}^{\nu} \left(
\varphi(\theta,X), \epsilon \varphi_{X}(\theta,X),
\, \dots \right) \, d^{m}\theta
\end{equation}
on the space of $2\pi$-periodic with respect to each
$\theta^{\alpha}$ functions $\varphi (\theta,X)$.

 Let us also introduce the functionals
\begin{equation}
\label{thf}
\theta_{0}^{* \alpha}(X) \,\,\, = \,\,\, 
\theta^{\alpha}_{0}(X) \,\,\, - \,\,\,
\theta^{\alpha}_{0}(X_{0}) \,\,\, - \,\,\, {1 \over \epsilon} 
\int_{X_{0}}^{X} k^{\alpha}(J(X^{\prime})) \,\, dX^{\prime}
\end{equation}
for some fixed point $X_{0}$. We have identically
\begin{equation}
\label{vtrestr}
\theta^{* \alpha}_{0}(X_{0}) \,\,\, \equiv \,\,\, 0
\end{equation}

 As was shown in \cite{engam}, we can
also obtain the values of $U^{\nu}(X)$ and 
$\theta^{\alpha}_{0}(X)$ from $J^{\nu}(X)$,
$\theta^{*\alpha}_{0}(X)$ and 
$\theta^{\alpha}_{0}(X_{0})$ on
${\cal M}^{\prime}$
as formal series in powers of
$\epsilon$ and we will have for these series
\begin{equation}
\label{form1}
U^{\nu}(X) [J,\theta_{0}^{*}] \,\,\, = \,\,\, 
J^{\nu}(X) \,\,\, + \,\,\, \sum_{k\geq 1} \,
\epsilon^{k} \, u^{\nu}_{(k)}(J,J_{X},\theta^{*}_{0X},\dots)
\end{equation}
\begin{equation}
\label{form2}
\theta^{\alpha}_{0}(X) [J,\theta_{0}^{*}] \,\,\, = \,\,\,
\theta^{* \alpha}_{0}(X) \,\,\, + \,\,\, 
\theta^{\alpha}_{0}(X_{0}) \,\,\, + \,\,\,
{1 \over \epsilon} 
\int_{X_{0}}^{X} k^{\alpha}(J(X^{\prime})) \,\, dX^{\prime}
\end{equation}

 The form of the relation (\ref{form1}) will be important
in our considerations, so we reproduce here the 
calculations from \cite{engam}.

 We remind that the values
$J^{\nu}(X)$, $\theta_{0}^{*\alpha}(X)$,
$\theta_{0}^{\alpha}(X_{0})$
and $U^{\mu}(X)$ are connected on ${\cal M}^{\prime}$ by the
relations (the definition of $J^{\nu}(X)$):
$$J^{\nu}(X) \,\,\, = \,\,\, 
{1 \over (2\pi)^{m}}\int_{0}^{2\pi}\!\!\!\dots\int_{0}^{2\pi}
{\cal P}^{\nu} \left(
\Phi_{(in)}(\theta + s(X,\epsilon), U(X)), \,
\epsilon \partial_{X} \Phi_
{(in)}(\theta + s(X,\epsilon), U(X)), \dots \right) 
\, d^{m}\theta \,\, =$$
$$= \,\,\, {1 \over (2\pi)^{m}}
\int_{0}^{2\pi}\!\!\!\dots\int_{0}^{2\pi}
{\cal P}^{\nu} \left(\Phi_{(in)}(\theta + s(X,\epsilon),U(X)), 
\, k^{\alpha}(J) \, \partial_{\theta^{\alpha}} \Phi_{(in)}
(\theta + s(X,\epsilon), U(X)),\dots \right) d^{m}\theta 
\,\,\, + $$
$$+ \,\,\, \sum_{k \geq 1} \, \epsilon^{k} \,
{1 \over (2\pi)^{m}}
\int_{0}^{2\pi}\!\!\!\dots\int_{0}^{2\pi} 
{\cal P}^{\nu}_ {(k)}
\left(\Phi_{(in)}(\theta + s(X,\epsilon), U(X)), \,
\dots \right) \, d^{m}\theta  \quad , $$
where 
$$s(X,\epsilon) \,\,\, \equiv \,\,\,
\theta^{*}_{0}(X) \,\,\, + \,\,\, \theta_{0}(X_{0}) 
\,\,\, + \,\,\, {1 \over \epsilon}\int_{X_{0}}^{X} k
(J(X^{\prime}))\, dX^{\prime} $$
and 
$\, {\cal P}^{\nu}_{(k)}(\Phi_{(in)}(\theta + s(X,\epsilon),\dots)$
are local densities depending on
$\, \Phi_{(in)}(\theta + s(X,\epsilon), U(X))$ and their 
derivatives with respect to
$U^{\nu}$ and $\theta^{\alpha}$ with the coefficients of type:
$\, U_{X}(X)$, $\, U_{XX}(X),\dots $,
$\, k(J)$, $\, \partial_{X} k(J)$, $\, \partial_{X}^{2} k(J) $,
$\, \dots$, and $\, \theta_{0X}^{*}(X) $, 
$\, \theta_{0XX}^{*}(X),\dots$,
given by the collecting together these terms, having the general
multiplier $\epsilon^{k}$. The term corresponding to the zero 
power of $\epsilon$ is written separately.

 After the integration with respect to $\theta$, which removes 
the singular at $\epsilon \rightarrow 0$ phase shift 
$\theta_{0}$ in the argument of $\Phi_{(in)}$,
we obtain on ${\cal M}^{\prime}$:
\begin{equation}
\label{cd}
J^{\nu}(X) \,\,\, = \,\,\,
\zeta^{\nu}(J,U) \,\,\, + \,\,\, \sum_{k \geq 1} \,
\epsilon^{k} \,
\zeta^{\nu}_{(k)} \left(
U,U_{X},\dots,U_{kX}, \, J,J_{X},\dots,J_{kX},
\, \theta_{0X}^{*},\dots, \theta_{0kX}^{*} \right) 
\end{equation}

 The sum in (\ref{cd}) contains a finite number of terms.
The functions $\, \zeta^{\nu}_{(k)}$ and $\, \zeta^{\nu}$ 
are the integrated with respect to
$\theta$ functions $\, {\cal P}^{\nu}_{(k)}$ and
$\, {\cal P}^{\nu}$ respectively.

 So, since
$$\zeta^{\nu}(J,U) \,\,\, = \,\,\, 
{1 \over (2\pi)^{m}}\!\int_{0}^{2\pi}\!\!\!\dots\int_{0}^{2\pi}
{\cal P}^{\nu} \left(\Phi_{(in)}(\theta,U), \,
k^{\alpha}(J) \, \Phi_{(in)\theta^{\alpha}}(\theta,U),
\, \dots \right) \, d^{m}\theta $$
we obtain that the system
\begin{equation}
\label{de}
J^{\nu}(X) \,\,\, = \,\,\, \zeta^{\nu} \big( J(X),U(X) \big)
\end{equation}
is satisfied by the solution $\, J^{\nu}(X) \equiv U^{\nu}(X)$
according to the definition of the parameters $U^{\nu}$.
Since we suppose that system (\ref{de})
has a generic form we will assume 
that (locally) this is the 
only solution and put $J^{\nu}(X) = U^{\nu}(X)$ in the zero
order of $\epsilon$.

 After that we can resolve system (\ref{cd}) by
iterations, taking on the initial step  \linebreak
$U^{\nu}(X) = J^{\nu}(X)$.
The substitution of (\ref{form1}) into (\ref{cd}) 
under the condition of the non-singularity of matrix
$\, \|{\partial \zeta^{\nu}(J,U) \over \partial U^{\mu}}\||_
{U = J} \, $  
will sequentially define the functions $u^{\nu}_{(k)}$.
So we obtain the relations (\ref{form1}) and (\ref{form2}).

\vspace{0.5cm}

 Now we can take  also the values of $\, J^{\nu}(X)$,
$\, \theta^{*\alpha}_{0}(X)$, $\, \theta^{\alpha}_{0}(X_{0})$
and 
$\, G^{i}_{[U[\varphi],\theta_{0}[\varphi]]} (\theta,X)$ 
with the restrictions (\ref{vtrestr}) and also
\begin{equation}
\label{res11}
{1 \over (2\pi)^{m}}\int_{0}^{2\pi}\!\!\!\dots\int_{0}^{2\pi}
\!\! {\tilde \kappa}_{(q)[U[\varphi](X)]}
\left(\theta + \theta_{0}^{*}(X) + \theta_{0}(X_{0}) +
{1 \over \epsilon}
\int_{X_{0}}^{X} \!\! k(J(X^{\prime})) \, dX^{\prime}\right) 
G^{i}_{[U[\varphi],\theta_{0}[\varphi]]}
(\theta,X) \, d^{m}\theta \,\,\, \equiv \,\,\, 0
\end{equation}
as coordinates in the vicinity of ${\cal M}^{\prime}$.

\vspace{0.5cm}

 We define now a Poisson bracket on the space of functions
$\varphi (\theta,X)$ by the formula
$$\{\varphi^{i}(\theta,X),\varphi^{j}(\theta^{\prime},Y)\} 
\,\,\, = \,\,\,
\sum_{k\geq 0} B^{ij}_{k}(\varphi, \epsilon\varphi_{X}, \dots)
\, \epsilon^{k} \, \delta^{(k)}(X - Y) \, 
\delta(\theta - \theta^{\prime}) \,\,\, +$$ 
\begin{equation}
\label{newbr}
+ \,\,\, {1 \over \epsilon} \,\, 
\delta(\theta - \theta^{\prime}) \, \sum_{k\geq 0} \, e_{k} \,
S^{i}_{(k)}(\varphi, \epsilon\varphi_{X}, \dots)
\,\, \nu (X - Y) \,\, 
S^{j}_{(k)}(\varphi, \epsilon\varphi_{Y}, \dots)
\end{equation}
which is just a rescaling of the bracket (\ref{canform}), 
multiplied by $\delta(\theta - \theta^{\prime})$.
We normalize here the $\delta$-function
$\delta(\theta - \theta^{\prime})$ by $(2\pi)^{m}$.
 
 The pairwise Poisson brackets of the constraints 
$G^{i}_{[U,\theta_{0}]}(\theta,X)$ on ${\cal M}^{\prime}$ can
be written in the form
\begin{multline}
\label{pbgi}
\{G^{i}(\theta,X), G^{j}(\theta^{\prime},Y)\}|_
{{\cal M}^{\prime}} \,\,\, = \,\,\, 
{1 \over (2\pi)^{2m}}\int_{0}^{2\pi}\!\!\dots\int_{0}^{2\pi}
L^{i}_{k [U(X),\theta_{0}(X)]}(\theta, \tau) \, \times  \\
\times \, L^{j}_{s [U(Y),\theta_{0}(Y)]}(\theta^{\prime}, \sigma)
\times \{\varphi^{k}(\tau,X), \varphi^{s}(\sigma,Y)\}
|_{{\cal M}^{\prime}} \, d^{m}\tau \, d^{m}\sigma
\end{multline}
(we can omit the Poisson brackets of $L^{i}_{k}$ and
$L^{j}_{s}$ on ${\cal M}^{\prime}$ and also the brackets of the
functionals $\theta^{\alpha}_{0}[\varphi]$ and
$U^{\nu}[\varphi]$ from $\Phi_{(in)}$ in (\ref{svgi})
since they are multiplied by the convolutions of the
corresponding $L$-operators with the ``right eigen vectors''
$\Phi_{(in)\theta^{\alpha}}$ and
$\Phi_{(in)U^{\nu}}$, which are zero on ${\cal M}^{\prime}$).

 Brackets (\ref{pbgi}) evidently satisfy 
the orthogonality conditions:
\begin{multline}
\label{uslovie1}
{1 \over (2\pi)^{m}}\int_{0}^{2\pi}\!\!\!\dots\int_{0}^{2\pi}
\!\!{\tilde \kappa}_{(q)i [U[J,\theta^{*}_{0}](X)]}
\left(\theta + \theta^{*}_{0}(X) + \theta_{0}(X_{0}) +
{1 \over \epsilon} 
\int_{X_{0}}^{X} k(J(X^{\prime})) dX^{\prime}\right)
\times  \\
\times \{G^{i}(\theta,X), G^{j}(\theta^{\prime},Y)\}|_
{{\cal M}^{\prime}} \, d^{m}\theta \,\,\, = \,\,\, 0
\end{multline}
\begin{multline}
\label{uslovie2}
{1 \over (2\pi)^{m}}\int_{0}^{2\pi}\!\!\dots\int_{0}^{2\pi}
\{G^{i}(\theta,X), G^{j}(\theta^{\prime},Y)\}|_
{{\cal M}^{\prime}} \, \times  \\
\times \, {\tilde \kappa}_{(q)j [U[J,\theta^{*}_{0}](Y)]}
\left(\theta + \theta^{*}_{0}(Y) + \theta_{0}(X_{0}) +
{1 \over \epsilon}
\int_{X_{0}}^{Y} k(J(Y^{\prime})) dY^{\prime}\right)
d^{m}\theta^{\prime} \,\,\, = \,\,\, 0
\end{multline}
for $\, q = 1,\dots,N+m$ in the coordinates
$\, J(X)$, $\, \theta^{*}_{0}(X)$ and $\, \theta_{0}(X_{0})$
on the submanifold $\, {\cal M}^{\prime}$.

 We note now that every derivative with respect to
$X$ or $Y$ appears in the bracket (\ref{newbr}) with
the multiplier $\epsilon$ but, being applied to the
functions

\begin{equation}
\label{funform}
\varphi^{i}(\theta,X) \,\,\, = \,\,\,  
\Phi^{i}_{(in)}\left(\theta + \theta^{*}_{0}(X) + 
\theta_{0}(X_{0}) + {1 \over \epsilon}
\int_{X_{0}}^{X} k(J(X^{\prime})) dX^{\prime},
U[J,\theta^{*}_{0}](X)\right)
\end{equation}
on ${\cal M}^{\prime}$, contains the nonzero at 
$\epsilon \rightarrow 0$ term 
$\, k^{\alpha}(J) \, \partial/\partial \theta^{\alpha}$. 
Now we formulate the statement about the structure of the
bracket (\ref{newbr}) on ${\cal M}^{\prime}$ in the coordinates
$\, J(X)$, $\, \theta^{*}_{0}(X)$ and $\, \theta_{0}(X_{0})$.

\vspace{0.5cm}

{\bf Lemma 2.3}

{\it
 The pairwise Poisson brackets of constraints 
$\, G^{i}_{[U,\theta_{0}]}(\theta, X)$ on ${\cal M}^{\prime}$
have no singular terms
at $\, \epsilon \rightarrow 0$ and no non-local terms
in the zero order of $\epsilon$ $(\epsilon^{0})$
at any fixed coordinates 
$\, J^{\nu}(X)$, $\, \theta^{*\alpha}_{0}(X)$
and $\, \theta^{\alpha}_{0}(X_{0})$
(such that $\, U(X) = U[J,\theta^{*}_{0}](X)$,
$\, \theta_{0}(X) = \theta_{0}[J,\theta^{*}_{0},
\theta_{0}(X_{0})](X)$). }

\vspace{0.5cm}

{\it Proof.}

 The first statement is evident for the local part of bracket
(\ref{newbr}), since any differentiation with respect to $X$
in it appears with the multiplier $\epsilon$ and has the
regular at $\epsilon \rightarrow 0$ form 
$k^{\alpha}(J(X))\partial/\partial \theta^{\alpha} + 
O(\epsilon)$, being applied to the functions of the form 
(\ref{funform}). So, we should check only the non-local part
of (\ref{newbr}), which contains the multiplier $\epsilon^{-1}$
in it. But, according to the relation (\ref{ssviaz}) and also
(\ref{form1}), we have that the terms arising on 
the both sides of $\, \nu (X-Y)$ (the convolutions of ${\hat L}$
with $\, S_{(k)}(\Phi,k^{\alpha}\Phi_{\theta^{\alpha}},\dots)$)
are of order of $\, \epsilon$ on $\, {\cal M}^{\prime}$
in the coordinates $\, J^{\nu}(X)$ and 
$\, \theta^{*\alpha}_{0}(X)$. So, we obtain that all the 
non-local part of (\ref{pbgi}) is of the order of $\epsilon$ on 
$\, {\cal M}^{\prime}$ at any fixed
coordinates $\, J^{\nu}(X)$, $\, \theta^{*\alpha}_{0}(X)$ and
$\, \theta^{\alpha}_{0}(X_{0})$.
 
{\hfill {\it Lemma 2.3 is proved.}}

\vspace{0.5cm}

 Let us formulate now the last ``regularity'' property of the
submanifold ${\cal M}^{\prime}$ with respect to the Poisson
structure (\ref{newbr}).

\vspace{0.5cm}

 We consider in the
coordinates $\, J^{\nu}(X)$, $\, \theta^{*\alpha}_{0}(X)$
and $\, \theta^{\alpha}_{0}(X_{0})$
on $\, {\cal M}^{\prime}$ a linear non-homogeneous system
on the functions

$$f_{j[J,\theta^{*}_{0}]}
\left(\theta^{\prime} + \theta^{*}_{0}(Y) + 
\theta_{0}(X_{0}) +
{1 \over \epsilon} \int_{X_{0}}^{Y} k(J(Y^{\prime})) 
dY^{\prime}, Y, \epsilon\right)  \quad , $$
having the form
\begin{multline*}
{1 \over (2\pi)^{m}}\int_{0}^{2\pi}\!\!\dots\int_{0}^{2\pi}
\{G^{i}_{[U[\varphi],\theta_{0}[\varphi]]}(\theta,X),
G^{j}_{[U[\varphi],\theta_{0}[\varphi]]}(\theta^{\prime},Y)\}
|_{{\cal M}^{\prime}} \, \times  \\
\times \, f_{j}\left(\theta^{\prime} + \theta^{*}_{0}(Y) + 
\theta_{0}(X_{0}) + {1 \over \epsilon}
\int_{X_{0}}^{Y} k(J(Y^{\prime})) dY^{\prime},
Y, \epsilon\right) d^{m}\theta^{\prime} dY \,\,\, =
\end{multline*}
\begin{equation}
\label{lins}
= \,\,\, \{G^{i}_{[U[\varphi],\theta_{0}[\varphi]]}(\theta,X),
F[\varphi](\epsilon)\}|_{{\cal M}^{\prime}}
\end{equation}
where $F[\varphi](\epsilon)$ is a functional, defined in the
vicinity of ${\cal M}^{\prime}$.

 After all differentiations with respect to $X$ we can omit
the term 
$$\theta^{*}_{0}(X) + \theta_{0}(X_{0}) + {1 \over \epsilon}
\int_{X_{0}}^{X} k^{\alpha}(J(X^{\prime})) dX^{\prime} 
\quad , $$
which appears in all functions depending on $\theta$ 
and $X$ in (\ref{lins}), and then consider the system 
(\ref{lins}) at the zero order of $\epsilon$.

 From Lemma 2.3 we have that in the zero order of $\epsilon$
the brackets 
$\, \{G^{i}(\theta,X),G^{j}(\theta^{\prime},Y)\}$
on ${\cal M}^{\prime}$ do not include non-local terms,
containing $\, \nu (X-Y)$. For the derivatives with respect
to $X$, which arise with the multiplier $\, \epsilon$
from the local terms of
$\, \{\varphi^{k}(\tau,X),\varphi^{s}(\sigma,Y)\}|_
{{\cal M}^{\prime}} \, $, we should take in the zero order of 
$\, \epsilon$ only the main part 
$\, k^{\alpha}(J) \, \partial/\partial \theta^{\alpha}$.
So, in the zero order of $\epsilon$ we obtain from 
(\ref{lins}) just linear systems of  
integro-differential equations with respect to $\, \theta$
and $\, \theta^{\prime}$ on the functions 
$\, f_{j}(\theta^{\prime},X)$, independent at different $X$.
We have also that the right-hand side of (\ref{lins}) 
satisfies at any $X$ and $\epsilon$ the compatibility 
conditions (\ref{uslovie1})  
 (let us remind that $\, U^{\nu}[J,\theta^{*}_{0}]$ 
are the asymptotic series at
$\, \epsilon \rightarrow 0$).

\vspace{0.5cm}

{\it  (IV) We require that the system 
(\ref{lins}) is resolvable on $\, {\cal M}^{\prime}$
for any $\, F[\varphi](\epsilon)$
in the class of $2\pi$-periodic with respect to all
$\theta^{\alpha}$ functions and its solutions can be
represented in the form of regular at 
$\epsilon \rightarrow 0$ asymptotic series
$$f_{j[J,\theta^{*}_{0}]}(\theta, Y, \epsilon) 
\,\,\, = \,\,\, \sum_{n\geq 0} \, \epsilon^{k} \,
f^{(k)}_{j[J,\theta^{*}_{0}]}(\theta, Y) $$
for regular at $\epsilon \rightarrow 0$ right-hand sides
of (\ref{lins}).}\footnote{Let us say that this requirement
is satisfied for a wide class of Poisson brackets
(\ref{canform}), however, it is not necessary fulfilled
in the general case. It can be actually shown that this
requirement can be significantly weakened and replaced
by resolvability of system (\ref{lins}) just for everywhere
dense set of parameters $\, U[\varphi]$ on 
$\, {\cal M}^{\prime}$, using the approach represented in
\cite{Sigma,MinimalSet}. We will, however, use here the 
assumption, formulated above, since the methods used in
\cite{Sigma,MinimalSet} require in fact noticeably longer
considerations.}

\vspace{0.5cm}

 The condition (IV) is responsible for the Dirac restriction
of the bracket (\ref{newbr}) on the submanifold 
${\cal M}^{\prime}$.

\vspace{0.5cm}

 Now we prove a statement which will be very important
for our averaging procedure.

\vspace{0.5cm}

{\bf Lemma 2.4}

{\it
 The Poisson brackets of the functionals 
$\theta^{*\alpha}_{0}(X)$ with
$J^{\nu}(Y)$  are of order of $\epsilon$ 
at $\epsilon \rightarrow 0$ on ${\cal M}^{\prime}$
at any fixed coordinates $\, J^{\nu}(X)$,
$\, \theta^{*\alpha}_{0}(X)$ and 
$\, \theta^{\alpha}_{0}(X_{0})$ : 
\begin{equation}
\label{jthskob}
\{\theta^{*\alpha}_{0}(X), J^{\nu}(Y)\}|_{{\cal M}^{\prime}} 
\,\,\, = \,\,\, O(\epsilon) \quad , \quad \quad 
\epsilon \rightarrow 0
\end{equation} 
}

\vspace{0.5cm}

{\it Proof.}

 First we note that the Poisson brackets of 
$\varphi^{i}(\theta,X)$
with the functionals $J^{\nu}(Y)$ can be written in the form
$$\{\varphi^{i}(\theta,X),J^{\nu}(Y)\} 
\,\,\, = \,\,\, \sum_{k\geq 0}
C^{i\nu}_{k} \left( \varphi(\theta,X),
\epsilon\varphi_{X}(\theta,X), \dots \right) 
\, \epsilon^{k} \, \delta^{(k)}(X - Y) \,\,\, + $$
$$+ \,\,\, \sum_{k\geq 0} \, e_{k} \,  
S^{i}_{(k)}(\varphi \left( \theta,X),
\epsilon\varphi_{X}(\theta,X),\dots \right) 
\, \nu(X-Y) \, \left(F^{\nu}_{(k)} (\varphi(\theta,Y),
\epsilon\varphi_{Y}(\theta,Y), \dots)
\right)_{Y} $$
for some $\, C^{i\nu}_{k}(\varphi,\epsilon\varphi_{X},\dots)$
and $\, F^{\nu}_{(k)}(\varphi,\epsilon\varphi_{Y},\dots)$ 
(we have integrated with respect to $\theta^{\prime}$).
 
 So, the flow generated by the functional 
$\int q(Y)J^{\nu}(Y)dY$ (where $q(Y)$ has a compact support) 
can be written as
\begin{multline*}
\varphi^{i}_{t} \,\,\, = \,\,\, \sum_{k\geq 0}
C^{i\nu}_{k}(\varphi,\epsilon\varphi_{X},\dots) \,
\epsilon^{k} q_{kX}(X) \,\,\, +  \\
+ \,\,\, \sum_{k\geq 0} \, e_{k} \,
S^{i}_{(k)}(\varphi,\epsilon\varphi_{X},\dots)
\int \nu(X-Y) \, q(Y) \, {d \over dY} 
F^{\nu}_{(k)}(\varphi,\epsilon\varphi_{Y},\dots) \, dY 
\,\,\, = 
\end{multline*}
\begin{multline}
\label{qintfl}
= \,\,\, \sum_{k\geq 0} 
C^{i\nu}_{k}(\varphi,\epsilon\varphi_{X},\dots) 
\epsilon^{k} q_{kX}(X) \,\,\, + \,\,\, \sum_{k\geq 0} e_{k} 
S^{i}_{(k)}(\varphi,\epsilon\varphi_{X},\dots) \,
F^{\nu}_{(k)}(\varphi,\epsilon\varphi_{X},\dots) \, 
q(X) \,\,\, -  \\
- \sum_{k\geq 0} \, e_{k} \, 
S^{i}_{(k)}(\varphi,\epsilon\varphi_{X},\dots)
\int \nu(X-Y) F^{\nu}_{(k)}(\varphi,\epsilon\varphi_{Y},\dots)
\, q_{Y}(Y) \, dY
\end{multline}

 As can be easily seen, the local terms of (\ref{qintfl}) have 
the form
$$q(X) \left[ C^{i\nu}_{0}(\varphi,\epsilon\varphi_{X},\dots) 
\, + \, \sum_{k\geq 0} e_{k} 
S^{i}_{(k)}(\varphi,\epsilon\varphi_{X},\dots)
F^{\nu}_{(k)}(\varphi,\epsilon\varphi_{X},\dots) \right] 
\,\,\, + \,\,\, O(\epsilon) $$
where the term in the brackets is just the flow, generated by
the functional
$${1 \over (2\pi)^{m}} \int 
\int_{0}^{2\pi}\!\!\dots\int_{0}^{2\pi} {\cal P}^{\nu}
(\varphi,\epsilon\varphi_{X},\dots) \, d^{m}\theta \, dX $$

In the non-local part of (\ref{qintfl}) (the last expression)
we have the convolution of the ``slow'' functions $q_{Y}(Y)$
with the rapidly oscillating 
$\, F^{\nu}_{(k)}(\varphi,\epsilon \varphi_{Y},\dots)$,
where $\, \varphi^{i}(\theta,Y)$ has the form (\ref{funform}).
So, in the leading order of $\epsilon$ we can neglect the 
dependence on $\theta$ of the last integral of (\ref{qintfl})
and take the averaged with respect to $\theta$ values 
$\, \langle F^{\nu}_{(k)} \rangle$ on ${\cal M}^{\prime}$
instead of the exact 
$\, F^{\nu}_{(k)}(\varphi,\epsilon \varphi_{Y},\dots)$
in the integral expression in (\ref{qintfl}). 

 After that we obtain, that the non-local term of 
(\ref{qintfl}) gives us in the zero order of $\epsilon$
a linear combination
of the flows $\, S_{(k)}(\varphi,\epsilon\varphi_{X},\dots)$,
considered on the functions
$$\varphi^{i}(\theta,X) \,\,\, = \,\,\, \Phi^{i}_{(in)}
\left(\theta + \theta^{*}_{0}(X) + \theta_{0}(X_{0}) +
{1 \over \epsilon}
\int_{X_{0}}^{X} k(J(X^{\prime})) \, dX^{\prime}, \, 
J(X)\right)  \quad , $$ 
at any fixed point $X$.

 From the invariance of the submanifold ${\cal M}$ with respect
to the flows (\ref{sperfl}) and (\ref{nuperfl}) we can conclude
now that the flow (\ref{qintfl}), being considered at the points 
of ${\cal M}^{\prime}$ with fixed coordinates $J(X)$,
$\theta^{*}_{0}(X)$, $\theta_{0}(X_{0})$ 
in the zero order of $\epsilon$, leaves ${\cal M}^{\prime}$
invariant and generates on it a linear evolution of the 
initial phases

$$\theta^{\alpha}_{0}(X) \,\,\, = \,\,\, 
\theta^{*\alpha}_{0}(X) + 
\theta_{0}^{\alpha}(X_{0}) + {1 \over \epsilon}
\int_{X_{0}}^{X} k^{\alpha}(J(X^{\prime})) \, dX^{\prime} $$
with some frequencies $\Omega^{\alpha\nu}_{[q]}(X)$ .
Here we use the formula (\ref{form1}) for
$U[J,\theta^{*}_{0}]$ and we can claim now that the Poisson
brackets of the functionals $\theta^{\alpha}_{0}(X)$ with
$\int q(Y)J^{\nu}(Y) dY$ at the points of ${\cal M}^{\prime}$
with fixed coordinates $J^{\nu}(X)$, 
$\theta^{*\alpha}_{0}(X)$ and $\theta_{0}(X_{0})$ have the 
form
\begin{equation}
\label{thqnubr}
\{\theta^{\alpha}_{0}(X), \int q(Y) \, J^{\nu}(Y) \, dY \} 
\,\,\, = \,\,\, 
\Omega^{\alpha\nu}_{[q]}[J,\theta^{*}_{0}](X) 
\,\,\, + \,\,\, O(\epsilon)
\end{equation}

 Let us now prove the relation
\begin{equation}
\label{kqnubr}
\{k^{\alpha}(J(X)), \int q(Y) \, J^{\nu}(Y) \, dY\} 
\,\,\, = \,\,\, \epsilon \,\, {d \over dX} \,\, 
\Omega^{\alpha\nu}_{[q]}[J,\theta^{*}_{0}](X) \,\,\,
+ \,\,\, O(\epsilon^{2})
\end{equation}
at the points of $\, {\cal M}^{\prime}$ with fixed values
of $\, J^{\nu}(X)$, $\, \theta^{*\alpha}_{0}(X)$ and
$\, \theta^{\alpha}_{0}(X_{0})$.

 Using again the relation (\ref{form1}) we can write for
(\ref{qintfl}) at the points of ${\cal M}^{\prime}$
$$\varphi^{i}_{t} \,\,\, = \,\,\, \Omega^{\beta\nu}_{[q]}(X)
\, \Phi^{i}_{(in)\theta^{\beta}}
\left(\!\theta + \theta^{*}_{0}(X) + \theta_{0}(X_{0}) +
{1 \over \epsilon}\!
\int_{X_{0}}^{X}\!\! k(J(X^{\prime})) dX^{\prime}, \, 
U[J,\theta^{*}_{0}](X)\!\right)\!\! \,\,\, + $$
\begin{equation}
\label{phit}
+ \,\,\, \epsilon \eta^{i}\left(\theta + \theta^{*}_{0}(X) +
\theta_{0}(X_{0}) + {1 \over \epsilon}
\int_{X_{0}}^{X} k(J(X^{\prime})) dX^{\prime}, \,
[J,\theta^{*}_{0}]\right)
\end{equation}
where $[J,\theta^{*}_{0}]$ means a regular at 
$\, \epsilon \rightarrow 0$
dependence on $\, J, \, J_{X}, \, \theta^{*}_{0X},\dots$.

 We are interested in the evolution of the functionals
$$J^{\mu}(X) \,\,\, = \,\,\, 
{1 \over (2\pi)^{m}}\int_{0}^{2\pi}\!\!\dots\int_{0}^{2\pi}
{\cal P}^{\mu} (\varphi,\epsilon\varphi_{X},\dots) \, 
d^{m}\theta $$

 We have
$${d \over dt} \, J^{\mu}(X) \,\,\, = \,\,\,
{1 \over (2\pi)^{m}}\int_{0}^{2\pi}\!\!\!\dots\int_{0}^{2\pi}
\!\! \left({\partial {\cal P}^{\mu} \over \partial \varphi^{i}}
\varphi^{i}_{t} + 
{\partial {\cal P}^{\mu} \over \partial \varphi^{i}_{X}}
\varphi^{i}_{tX} +
{\partial {\cal P}^{\mu} \over \partial \varphi^{i}_{XX}}
\varphi^{i}_{tXX} + \dots \right) d^{m}\theta \,\,\, = $$
$$= \,\,\,
{1 \over (2\pi)^{m}}\int_{0}^{2\pi}\!\!\dots\int_{0}^{2\pi}
\left( \Pi^{\mu}_{i(0)} \varphi^{i}_{t} + \Pi^{\mu}_{i(1)}
\epsilon \varphi^{i}_{tX} + \Pi^{\mu}_{i(2)}
\epsilon^{2} \varphi^{i}_{tXX} + \dots \right) d^{m}\theta$$
where the densities $\Pi^{\mu}_{i(k)}$ were introduced in
(\ref{adden}). 

 It is easy to see that (\ref{phit}) does not change $J^{\mu}(X)$
at the zero order of $\epsilon$ and we can also state that the 
terms of the order of $\epsilon$ in (\ref{phit}) (i.e. 
$\epsilon \eta^{i}(\theta + \dots , X)$) are unessential
for the evolution of $k(J(X))$ on ${\cal M}^{\prime}$ at the
order of $\epsilon$. Indeed, their contribution to the evolution
of $J^{\mu}(X)$ in the order of $\epsilon$ is
\begin{equation}
\label{dtjmu}
\epsilon \,
{1 \over (2\pi)^{m}}\int_{0}^{2\pi}\!\!\dots\int_{0}^{2\pi}
\left( \Pi^{\mu}_{i(0)} \eta^{i} + \Pi^{\mu}_{i(1)}
\epsilon \eta^{i}_{X} + \Pi^{\mu}_{i(2)}
\epsilon^{2} \eta^{i}_{XX} + \dots \right) d^{m}\theta 
\end{equation}
where we should take only the main term 
$\, k^{\gamma}(J(X)) \, \partial/\partial\theta^{\gamma}$
for the derivatives $\, \epsilon \, \partial/\partial X$ in the
formula (\ref{dtjmu}). After the integration by parts we have
for this contribution
$$\epsilon \,
{1 \over (2\pi)^{m}}\int_{0}^{2\pi}\!\!\dots\int_{0}^{2\pi}
\left( \Pi^{\mu}_{i(0)} - 
k^{\gamma}{\partial \over \partial\theta^{\gamma}}
\Pi^{\mu}_{i(1)} + \dots \right) \eta^{i}(\theta + \dots, X)
\, d^{m}\theta $$

 But after the substitution of the main part of 
$\, \varphi^{i}(\theta,X)$
$$\Phi^{i}_{(in)}
\left( \theta + \theta^{*}_{0}(X) + \theta_{0}(X_{0}) +
{1 \over \epsilon}
\int_{X_{0}}^{X} k(J(X^{\prime})) dX^{\prime},
\, J(X)\right) $$
(according to (\ref{form1})) into the densities 
$\Pi^{\mu}_{i(k)}(\varphi,\epsilon\varphi_{X},\dots)$ we 
obtain in the leading order of $\epsilon$ the convolution of 
$\eta(\theta,X)$ with the variational derivative of the
functional ${\bar I}^{\mu}$, introduced in (\ref{perfunnu}),
with respect to $\varphi(\theta,X)$. Our statement follows now
from Lemma 2.1, which claims that the variational derivatives
of the functionals $k^{\alpha}({\bar I}[\varphi])$ are
identically equal to zero on the space of $m$-phase solutions 
of (\ref{locsys}).

 Consider now the first term of (\ref{phit}). We have
that the evolution of $J^{\mu}(X)$, which is responsible
for the evolution of $k(J)$, is 
given by the expression
\begin{multline*}
{d \over dt} \, J^{\mu}(X) \,\,\,\,\, = \,\,\,\,\, 
\Omega^{\beta\nu}_{[q]}(X) \,\,
{1 \over (2\pi)^{m}}\int_{0}^{2\pi}\!\!\dots\int_{0}^{2\pi} 
\left( {\partial {\cal P}^{\mu} \over \partial \varphi^{i}} \,
\Phi^{i}_{(in)\theta^{\beta}}(\theta + s(X,\epsilon),
\, U[J,\theta^{*}_{0}](X)) \right. \, +  \\ 
+ \, \left. 
{\partial {\cal P}^{\mu} \over \partial \varphi^{i}_{X}} \,
{\partial \over \partial X} 
\Phi^{i}_{(in)\theta^{\beta}}(\theta + s(X,\epsilon),
\, U[J,\theta^{*}_{0}](X)) \, +
\dots \right) \, d^{m}\theta \,\,\, + 
\end{multline*}
\begin{multline*}
+ \,\,\, \epsilon\! \left(\Omega^{\beta\nu}_{[q]}(X)\right)_{X} 
{1 \over (2\pi)^{m}}\!
\int_{0}^{2\pi}\!\!\!\!\!\!\dots\int_{0}^{2\pi}
\!\!\left({\partial {\cal P}^{\mu} \over \partial \varphi^{i}_{X}}
+ 2 {\partial {\cal P}^{\mu} \over \partial \varphi^{i}_{XX}}
{\partial \over \partial X}
+ 3 {\partial {\cal P}^{\mu} \over \partial \varphi^{i}_{XXX}}
{\partial^{2} \over \partial X^{2}}
+ \dots\! \right) \times  \\
\times \, \Phi^{i}_{(in)\theta^{\beta}}
\left( \theta + s(X,\epsilon),
\, U[J,\theta^{*}_{0}](X) \right)
d^{m}\theta \,\,\, + \,\,\, O(\epsilon^{2})  \quad ,  
\end{multline*}
where $\, s(X,\epsilon) \, \equiv \, \theta^{*}_{0}(X) + 
\theta_{0}(X_{0}) + {1 \over \epsilon}
\int_{X_{0}}^{X} k(J(X^{\prime})) dX^{\prime}$ .

 The first term here after the substitution of exact 
$\varphi^{i}$ in the form
$$ \varphi^{i}(\theta,X) \,\,\, = \,\,\, 
\Phi^{i}_{(in)\theta^{\beta}}(\theta + s(X,\epsilon),
\, U[J,\theta^{*}_{0}](X))$$
on $\, {\cal M}^{\prime}$, as can be easily seen, is just

$$\Omega^{\beta\nu}_{[q]}(X) \,\, 
{1 \over (2\pi)^{m}}\int_{0}^{2\pi}\!\!\dots\int_{0}^{2\pi}
{\partial \over \partial\theta^{\beta}} \,
{\cal P}^{\mu}\left(\Phi_{(in)}(\dots),\Phi_{(in)X}(\dots),
\dots)\right) d^{m}\theta \,\,\, \equiv \,\,\, 0 \,\,\, , $$
while the second term on ${\cal M}^{\prime}$ in the leading 
order of $\epsilon$ is equal to
$$\epsilon \left(\Omega^{\beta\nu}_{[q]}(X)\right)_{X}
{1 \over (2\pi)^{m}}\int_{0}^{2\pi}\!\!\dots\int_{0}^{2\pi}
\sum_{p\geq 1} p \, \times $$
$$\times \, \Pi^{\mu}_{i(p)}\!\left(\!
\Phi_{(in)}(\theta + s(X,\epsilon),J(X)), \, 
k^{\gamma}(J(X)){\partial \over \partial\theta^{\gamma}}
\Phi_{(in)}(\theta + s(X,\epsilon),J(X)), \dots\!\right)
\times $$
\begin{equation}
\label{essev}
\times \, k^{\alpha_{1}}(J(X)) \dots k^{\alpha_{p-1}}(J(X))
\,\, \Phi^{i}_{(in)\theta^{\beta}
\theta^{\alpha_{1}}\dots\theta^{\alpha_{p-1}} }
\left(\theta + s(X,\epsilon),J(X)), \dots\right)
d^{m}\theta  \quad  ,
\end{equation}
which coincides with the integral expression from 
(\ref{kuprel}) in Lemma 2.2. 
So, from Lemma 2.2 we have that the summation of 
(\ref{essev}) with 
$\, \partial k^{\alpha}/\partial J^{\mu}$ is equal to
$\, \epsilon \left(\Omega^{\beta\nu}_{[q]}(X)\right)_{X} 
\delta^{\alpha}_{\beta} \, $ and we obtain that
$${\partial \over \partial t} k^{\alpha}(J) \,\,\ = \,\,\, 
\epsilon \, {\partial \over \partial X} \,
\Omega^{\alpha\nu}_{[q]}(X) \, + \, O(\epsilon^{2}) 
\quad , $$
i.e. the relation (\ref{kqnubr}).

 Now, using (\ref{thqnubr}) and (\ref{kqnubr}), we can write
that
$$\{\theta^{*\alpha}_{0}(X), \int q(Y)J^{\nu}(Y)dY\} 
\,\,\, = \,\,\,
\{\theta^{\alpha}_{0}(X) - \theta^{\alpha}_{0}(X_{0}) -
{1 \over \epsilon}
\int_{X_{0}}^{X} k^{\alpha}(J(X^{\prime})) dX^{\prime},
\int q(Y)J^{\nu}dY\} \,\,\, = \,\,\, O(\epsilon) $$
for any $q(Y)$ in our coordinates on ${\cal M}^{\prime}$ .

 So, we have 
$$\{\theta^{*\alpha}_{0}(X), 
J^{\nu}(Y)\}|_{{\cal M}^{\prime}} \,\,\, = \,\,\,
O(\epsilon) $$
at any fixed coordinates $J^{\nu}(X)$,
$\theta^{*\alpha}_{0}(X)$ and
$\theta^{\alpha}_{0}(X_{0})$.
 
{\hfill {\it Lemma 2.4 is proved.}}

\vspace{1cm}

\section{\bf Averaging procedure.}
\setcounter{equation}{0}

 Let us now describe the averaging procedure of the
Poisson bracket (\ref{canform}) on the family of $m$-phase 
solutions of (\ref{locsys}) under the conditions of 
``regularity'' formulated above.

\vspace{0.5cm}

{\bf Theorem 3.1}

{\it
 Let us have a Poisson bracket (\ref{canform}) and a local 
system (\ref{locsys}) generated by a local Hamiltonian
function

$$H \,\,\, = \,\,\, 
\int {\cal P}_{H}(\varphi,\varphi_{x},\dots) \, dx  \quad , $$
which has $N(\geq 2m)$-parametric full family of $m$-phase
solutions modulo $m$ initial phase shifts $\theta^{\alpha}_{0}$.

 Let us have $N$ commuting local translationally 
invariant integrals

$$I^{\nu} \,\,\, = \,\,\, 
\int {\cal P}^{\nu}(\varphi,\varphi_{x},\dots) \, dx $$
$$\{I^{\nu}, H\} \,\,\, = \,\,\, 0  \quad ,  
\quad \quad \,\,\, \{I^{\nu}, I^{\mu}\} \,\,\, = \,\,\, 0 
\quad , $$
which generate local flows according to
the Poisson bracket (\ref{canform})
and the averaged densities of which can be taken as parameters
$U^{\nu}$ on the space of $m$-phase solutions of (\ref{locsys})
(the conditions (A)-(C)).

 Then under the ``regularity'' conditions (I)-(IV) for the space of
$m$-phase solutions of (\ref{locsys}) we can construct a Poisson
bracket of Ferapontov type (\ref{ferbr}) for the ``slow''
parameters $U^{\nu}(X)$ by the following procedure:

 We calculate the pairwise Poisson brackets of
$\, {\cal P}^{\nu}(\varphi,\varphi_{x},\dots)$ in the form

$$\{{\cal P}^{\nu}(\varphi,\varphi_{x},\dots),
{\cal P}^{\mu}(\varphi,\varphi_{y},\dots)\} \,\,\, = \,\,\,
\sum_{k\geq 0} \, A^{\nu\mu}_{k}(\varphi,\varphi_{x},\dots) \,
\delta^{(k)}(x-y) \,\,\, + $$
$$+ \,\,\, \sum_{k\geq 0} e_{k}
\left(F^{\nu}_{(k)}(\varphi,\varphi_{x},\dots)\right)_{x}
\, \nu (x-y) \,
\left(F^{\mu}_{(k)}(\varphi,\varphi_{y},\dots)\right)_{y} $$
(where is a finite number of terms in the both sums). Here we
have the total derivatives of the functions 
$\, F^{\nu}_{(k)}$ and $\, F^{\mu}_{(k)}$ with respect 
to $x$ and $y$ as a corollary of the fact that both 
$\, I^{\nu}$ and $\, I^{\mu}$
generate local flows according to the Poisson
bracket (\ref{canform}). From the commutativity
of the set $\, \{I^{\nu}\}$ we have also

\begin{equation}
\label{vazhno}
A^{\nu\mu}_{0}(\varphi,\varphi_{x},\dots) \,\,\, + \,\,\,
\sum_{k\geq 0} e_{k}
\left(F^{\nu}_{(k)}(\varphi,\varphi_{x},\dots)\right)_{x}
F^{\mu}_{(k)}(\varphi,\varphi_{x},\dots) \,\,\, \equiv \,\,\,
\partial_{x} Q^{\nu\mu}(\varphi,\varphi_{x},\dots) 
\end{equation}
for some functions $\, Q^{\nu\mu}(\varphi,\varphi_{x},\dots)$.

 Then for the ``slow'' coordinates 
$\, U^{\nu}(X) = \langle{\cal P}^{\nu}\rangle(X)$ 
we can define a Poisson bracket by the formula

$$\{U^{\nu}(X), U^{\mu}(Y)\} \,\,\, = $$
$$= \,\,\, \left[\langle A^{\nu\mu}_{1}\rangle(X) \, + \,
\sum_{k\geq 0} e_{k}
\left(\langle F^{\nu}_{(k)}F^{\mu}_{(k)}\rangle -
\langle F^{\nu}_{(k)}\rangle \langle F^{\mu}_{(k)}\rangle 
\right)(X) \right] \, \delta^{\prime}(X-Y) 
\,\,\, +$$
$$+ \,\,\, \left[
{\partial \langle Q^{\nu\mu}\rangle(X) \over \partial X}  
\, - \, \sum_{k\geq 0} \, e_{k} \,
{\partial \langle F^{\nu}_{(k)}\rangle(X) \over \partial X}
\, \langle F^{\mu}_{(k)}\rangle(X) \right] \, \delta (X-Y) 
\,\,\, + $$
\begin{equation}
\label{usrbr}
+ \,\,\, \sum_{k\geq 0} \, e_{k} \, 
{\partial \langle F^{\nu}_{(k)}\rangle(X) \over \partial X}
\,\, \nu (X-Y) \,
{\partial \langle F^{\mu}_{(k)}\rangle(Y) \over \partial Y}
\quad  , 
\end{equation}
where the averaged values are the functions of $\, U(X)$ and
$\, U(Y)$ at the corresponding points $\, X$ and $\, Y$.

 Bracket (\ref{usrbr}) satisfies the Jacobi identity and
is invariant with respect to the choice of the set
$\, \{I^{1},\dots,I^{N}\}$, satisfying (A)-(C), if the choice 
of these integrals is not unique, i.e.

\vspace{2mm}

\noindent
if $\, U^{\nu} = \langle {\cal P}^{\nu}\rangle$,
$\, {\tilde U}^{\nu} = \langle {\tilde {\cal P}}^{\nu}\rangle$
and $\, \{U^{\nu}(X), U^{\mu}(Y)\}$, 
$\, \{{\tilde U}^{\nu}(X), {\tilde U}^{\mu}(Y)\}^{\prime}$
are the brackets (\ref{usrbr}), constructed with the aid of 
the sets $\{I^{\nu}\}$ and $\{{\tilde I}^{\nu}\}$ respectively, 
then
$$\{{\tilde U}^{\nu}(X), {\tilde U}^{\mu}(Y)\}^{\prime}
\,\,\, \equiv \,\,\,
{\partial {\tilde U}^{\nu} \over \partial U^{\lambda}}(X) \,\,
\{U^{\lambda}(X), U^{\sigma}(Y)\} \,\,
{\partial {\tilde U}^{\mu} \over \partial U^{\sigma}}(Y) $$

}

\vspace{0.5cm}

{\it Proof.}

 The most difficult part is to prove the Jacobi identity for the
bracket (\ref{usrbr}). For this we use the Dirac restriction
of the Poisson bracket (\ref{newbr}) on the submanifold
${\cal M}^{\prime}$ with the coordinates $J^{\nu}(X)$,
$\theta^{*\alpha}_{0}(X)$ and $\theta^{\alpha}_{0}(X_{0})$
on it. According to the Dirac
restriction procedure we should find for $J^{\nu}(X)$,
$\theta^{*\alpha}_{0}(X)$ 
and $\theta^{\alpha}_{0}(X_{0})$ the corrections of the form

$$V^{\nu}(X) \,\,\, = \,\,\,  
{1 \over (2\pi)^{m}}\int_{0}^{2\pi}\!\!\!\!\!\dots\int_{0}^{2\pi}
\int v^{\nu}_{j}[J,\theta^{*\alpha}_{0},\theta_{0}(X_{0})]
(X,\theta^{\prime},Z) \,\, G^{j}(\theta^{\prime},Z) \,\, 
d^{m}\theta^{\prime} \, dZ  \quad , $$

$$W^{\alpha}(X) \,\,\, = \,\,\, 
{1 \over (2\pi)^{m}}\int_{0}^{2\pi}\!\!\!\!\!\dots\int_{0}^{2\pi}
\int w^{\alpha}_{j}[J,\theta^{*\alpha}_{0},\theta_{0}(X_{0})]
(X,\theta^{\prime},Z) \,\, G^{j}(\theta^{\prime},Z) \,\, 
d^{m}\theta^{\prime} \, dZ  \quad , $$
and 

$$O^{\alpha} \,\,\, = \,\,\, 
{1 \over (2\pi)^{m}}\int_{0}^{2\pi}\!\!\!\!\!\dots\int_{0}^{2\pi}
\int o^{\alpha}_{j}[J,\theta^{*\alpha}_{0},\theta_{0}(X_{0})]
(\theta^{\prime},Z) \,\, G^{j}(\theta^{\prime},Z) \,\, 
d^{m}\theta^{\prime} \, dZ  \quad , $$
such that the fluxes, generated in the Hamiltonian structure 
(\ref{newbr}) by the``functionals''  \linebreak
$\, J^{\nu}(X) + V^{\nu}(X)$, 
$\, \theta^{*\alpha}_{0}(X) +
W^{\alpha}(X)$ and $\, \theta^{\alpha}_{0}(X_{0}) + O^{\alpha}$,
leave $\, {\cal M}^{\prime}$ invariant, i.e.

\begin{equation}
\label{uslvg}
{1 \over (2\pi)^{m}}\int_{0}^{2\pi}\!\!\dots\int_{0}^{2\pi}
\{G^{i}(\theta,Y), G^{j}(\theta^{\prime},Z)\}|_{{\cal M}^{\prime}} 
\times v^{\nu}_{j}(X,\theta^{\prime},Z)
\,\, d^{m}\theta^{\prime} \, dZ \,\,\, = \,\,\,  
- \, \{G^{i}(\theta,Y), J^{\nu}(X)\}|_{{\cal M}^{\prime}} 
\end{equation}
\begin{equation}
\label{uslwg}
{1 \over (2\pi)^{m}}\int_{0}^{2\pi}\!\!\dots\int_{0}^{2\pi}
\{G^{i}(\theta,Y), G^{j}(\theta^{\prime},Z)\}|_{{\cal M}^{\prime}}
\times w^{\alpha}_{j}(X,\theta^{\prime},Z)
\,\, d^{m}\theta^{\prime} \, dZ \,\,\, = \,\,\,
- \, \{G^{i}(\theta,Y), \theta^{*\alpha}_{0}(X)\}|_
{{\cal M}^{\prime}} 
\end{equation}
\begin{equation}
\label{uslog}
{1 \over (2\pi)^{m}}\int_{0}^{2\pi}\!\!\dots\int_{0}^{2\pi}
\{G^{i}(\theta,Y), G^{j}(\theta^{\prime},Z)\}|_{{\cal M}^{\prime}}
\times o^{\alpha}_{j}(\theta^{\prime},Z)
\,\, d^{m}\theta^{\prime} \, dZ \,\,\, = \,\,\, 
- \, \{G^{i}(\theta,Y), \theta^{\alpha}_{0}(X_{0})\}|_
{{\cal M}^{\prime}} 
\end{equation}

 After that we put for the Dirac restriction on 
$\, {\cal M}^{\prime}$
$$\{J^{\nu}(X), \, J^{\mu}(Y)\}_{D} \,\,\, = \,\,\,  
\{J^{\nu}(X) + V^{\nu}(X), J^{\mu}(Y) + V^{\mu}(Y)\}|_
{{\cal M}^{\prime}} \,\,\, = \,\,\,
\{J^{\nu}(X),J^{\mu}(Y)\}|_{{\cal M}^{\prime}} \,\,\, - $$
\begin{equation}
\label{jjrest}
- \,\,\, 
{1 \over (2\pi)^{2m}}\int_{0}^{2\pi}\!\!\dots\int_{0}^{2\pi}
v^{\nu}_{i}(X,\theta,Z) \times
v^{\mu}_{j}(Y,\theta^{\prime},Z^{\prime}) \times 
\{G^{i}(\theta,Z), G^{j}(\theta^{\prime},Z^{\prime})\}|_
{{\cal M}^{\prime}} \,\,\, d^{m}\theta \, d^{m}\theta^{\prime} \, 
dZ \, dZ^{\prime} 
\end{equation}
and, in the same way,

$$\{J^{\nu}(X),\theta^{*\alpha}_{0}(Y)\}_{D} \,\,\, = \,\,\,
\{J^{\nu}(X),\theta^{*\alpha}_{0}(Y)\}|_{{\cal M}^{\prime}} 
\,\,\, - $$
\begin{equation}
\label{jtrest}
- \,\,\,
{1 \over (2\pi)^{2m}}\int_{0}^{2\pi}\!\!\dots\int_{0}^{2\pi}
v^{\nu}_{i}(X,\theta,Z) \times
w^{\alpha}_{j}(Y,\theta^{\prime},Z^{\prime}) \times 
\{G^{i}(\theta,Z), G^{j}(\theta^{\prime},Z^{\prime})\}|_
{{\cal M}^{\prime}} \,\,\, d^{m}\theta \, d^{m}\theta^{\prime} \,
\,dZ \, dZ^{\prime} 
\end{equation}

$$\{\theta^{*\alpha}_{0}(X),\theta^{*\beta}_{0}(Y)\}_{D} 
\,\,\, = \,\,\,
\{\theta^{*\alpha}_{0}(X),\theta^{*\beta}_{0}(Y)\}|_
{{\cal M}^{\prime}} \,\,\, - $$
\begin{equation}
\label{ttrest}
- \,\,\, 
{1 \over (2\pi)^{2m}}\int_{0}^{2\pi}\!\!\dots\int_{0}^{2\pi}
w^{\alpha}_{j}(X,\theta,Z) \times
w^{\beta}_{j}(Y,\theta^{\prime},Z^{\prime}) \times 
\{G^{i}(\theta,Z), G^{j}(\theta^{\prime},Z^{\prime})\}|_
{{\cal M}^{\prime}} \,\,\, d^{m}\theta \, d^{m}\theta^{\prime} \,
\, dZ \, dZ^{\prime} 
\end{equation}
(and so on).

 After calculation of the brackets 
in (\ref{uslvg})-(\ref{uslog}) and the substitution of
$\varphi(\theta,X)$ in the form (\ref{funform}) we obtain
regular at $\epsilon \rightarrow 0$ systems
for the functions $\, {\bar v}^{\nu}_{j}(X,\theta,Z,\epsilon)$,
$\, {\bar w}^{\alpha}_{j}(X,\theta,Z,\epsilon)$ and
$\, {\bar o}^{\alpha}_{j}(\theta,Z,\epsilon)$, such that

$$v^{\nu}_{j}(X,\theta,Z,\epsilon) \,\,\, = \,\,\,
{\bar v}^{\nu}_{j}\left(X,\theta + \theta^{*}_{0}(Z) +
\theta_{0}(X_{0}) + {1 \over \epsilon}
\int_{X_{0}}^{Z} k(J(Z^{\prime})) dZ^{\prime}, \,
\epsilon\right) $$

$$w^{\alpha}_{j}(X,\theta,Z,\epsilon) \,\,\, = \,\,\, 
{\bar w}^{\alpha}_{j}\left(X,\theta + \theta^{*}_{0}(Z) +
\theta_{0}(X_{0}) + {1 \over \epsilon}
\int_{X_{0}}^{Z} k(J(Z^{\prime})) dZ^{\prime}, \,
\epsilon\right) $$
and

$$o^{\alpha}_{j}(\theta,Z,\epsilon) \,\,\, = \,\,\, 
{\bar o}^{\alpha}_{j}\left(\theta + \theta^{*}_{0}(Z) +
\theta_{0}(X_{0}) + {1 \over \epsilon}
\int_{X_{0}}^{Z} k(J(Z^{\prime})) dZ^{\prime}, \,
\epsilon\right)  \quad , $$
which coincide with the system (\ref{lins}).

 From the arguments analogous to those used in Lemma 2.3
and the fact that the flows, generated by the functionals 
$J^{\nu}(X)$, leave invariant the submanifold 
${\cal M}^{\prime}$ at the zero order of $\epsilon$
(at fixed coordinates $J(X)$, $\theta^{*}_{0}(X)$,
$\theta_{0}(X_{0})$) we have also that the right-hand
sides of these systems are regular at 
$\epsilon \rightarrow 0$ in these coordinates.

 So, according to (IV), we can find the functions
${\bar v}^{\nu}_{j}$, ${\bar w}^{\alpha}_{j}$ and
${\bar o}^{\alpha}_{j}$ in the form of regular at
$\epsilon \rightarrow 0$ asymptotic series.
(The functions  $v^{\nu}_{j}(X,\theta,Z,\epsilon)$,
$w^{\alpha}_{j}(X,\theta,Z,\epsilon)$ and 
$o^{\alpha}_{j}(\theta,Z,\epsilon)$ are not uniquely
defined but it is easy to show that this does affect 
the Dirac restriction of the bracket
(\ref{newbr}) on ${\cal M}^{\prime}$ according to the formulas
(\ref{jjrest})-(\ref{ttrest})).

 Besides that, as was mentioned above, the flows 
(\ref{qintfl}), generated by the functionals  \linebreak
$\int q(X) \, J^{\mu}(X) \, dX$ on the functions 
(\ref{funform}), leave invariant the submanifold
${\cal M}^{\prime}$ at the zero order of $\epsilon$ and generate
a linear evolution of the initial phases. So, we can conclude
that the right-hand side of the linear system (\ref{uslvg})
contains no zero powers of $\epsilon$ and we should 
start the expansion for 
${\bar v}^{\nu}_{i}(X,\theta,Z,\epsilon)$
from the first power. 
 
 Now we have

$${\bar v}^{\nu}_{j}[J,\theta^{*}_{0},\theta_{0}(X_{0})]
(X,\theta,Z,\epsilon) \,\,\, = \,\,\, 
\sum_{k\geq 1} \, \epsilon^{k} \,  
{\bar v}^{\nu}_{j (k)}[J,\theta^{*}_{0},\theta_{0}(X_{0})]
(X,\theta,Z) $$

$${\bar w}^{\alpha}_{j}[J,\theta^{*}_{0},\theta_{0}(X_{0})]
(X,\theta,Z,\epsilon) \,\,\, = \,\,\,   
\sum_{k\geq 0} \, \epsilon^{k} \, 
{\bar w}^{\alpha}_{j (k)}[J,\theta^{*}_{0},\theta_{0}(X_{0})]
(X,\theta,Z) $$

$${\bar o}^{\alpha}_{j}[J,\theta^{*}_{0},\theta_{0}(X_{0})]
(\theta,Z,\epsilon) \,\,\, = \,\,\,
\sum_{k\geq 0} \, \epsilon^{k} \,
{\bar o}^{\alpha}_{j (k)}[J,\theta^{*}_{0},\theta_{0}(X_{0})]
(\theta,Z) $$

According to the relations above and Lemma 2.3
we can see now that the corrections to the values
$\{J^{\nu}(X),J^{\mu}(Y)\}|_{{\cal M}^{\prime}}$ and
$\{\theta^{*\alpha}_{0}(X),J^{\mu}(Y)\}|_{{\cal M}^{\prime}}$
due to the Dirac restriction on ${\cal M}^{\prime}$
are of order of $O(\epsilon^{2})$
and $\, O(\epsilon)$ respectively.
 
 According to the relation (\ref{form1}) we can also substitute
the values $\, J^{\nu}(X)$ instead of 
$\, U^{\nu}[J,\theta^{*}_{0}](X)$ in the leading order of 
$\epsilon$ as the arguments of the averaged functions on 
$\, {\cal M}^{\prime}$.

 Now we calculate the values 
$\, \{J^{\nu}(X),J^{\mu}(Y)\}|_{{\cal M}^{\prime}}$.

 Let us consider the Poisson brackets

\begin{equation}
\label{qqjjbr}
\{\int q^{\nu}(X)J^{\nu}(X)dX,\int q^{\mu}(Y)J^{\mu}(Y)dY \}|_
{{\cal M}^{\prime}}
\end{equation}
for arbitrary smooth $q^{\nu}(X)$ and $q^{\mu}(Y)$ with 
compact supports. For the Poisson brackets of the densities
${\cal P}^{\nu}(\varphi,\epsilon \varphi_{X},\dots)$ 
according to (\ref{newbr}) we have the expression:

$$\{{\cal P}^{\nu}(\theta,X),{\cal P}^{\mu}(\theta^{\prime},Y)\}
\,\,\, = \,\,\, \sum_{k\geq 0} \, 
A^{\nu\mu}_{k}(\varphi,\epsilon \varphi_{X},\dots) \,\, 
\epsilon^{k} \, 
\delta^{(k)}(X-Y) \, \delta (\theta-\theta^{\prime}) 
\,\,\, + $$
$$+ \,\,\,
\epsilon \, \delta (\theta-\theta^{\prime}) \, \sum_{k\geq 0} 
\, e_{k} \,
\Big( F^{\nu}_{(k)}(\varphi,\epsilon\varphi_{X},\dots) \Big)_{X}
\,\, \nu (X-Y) \, 
\Big( F^{\mu}_{(k)}(\varphi,\epsilon\varphi_{Y},\dots) \Big)_{Y}
$$
such that

$$\{J^{\nu}(X),J^{\mu}(Y)\} \,\,\, = \,\,\,  
\sum_{k\geq 0} \epsilon^{k}
{1 \over (2\pi)^{m}}\int_{0}^{2\pi}\!\!\!\!\!\dots\int_{0}^{2\pi}
A^{\nu\mu}_{k}(\varphi(\theta,X),
\epsilon \varphi_{X}(\theta,X),\dots) \,\,
d^{m}\theta \,\,\, \delta^{(k)}(X-Y) \,\,\, + $$
\begin{multline}
\label{jnujmubr}
+ \,\,\, \epsilon \,\, \sum_{k\geq 0} \, e_{k} \, 
{1 \over (2\pi)^{m}}\int_{0}^{2\pi}\!\!\!\!\!\dots\int_{0}^{2\pi}
\Big( F^{\nu}_{(k)}
(\varphi(\theta,X), \,
\epsilon\varphi_{X}(\theta,X),\dots) \Big)_{X} \, 
\times  \\
\times \, \nu (X-Y) \,\,
\Big( F^{\mu}_{(k)}
(\varphi(\theta,Y), \,
\epsilon\varphi_{Y}(\theta,Y),\dots) \Big)_{Y} 
\,\,\, d^{m}\theta
\end{multline}

 Now we should substitute the functions 
$\, \varphi^{i}(\theta,X)$, $\, \varphi^{i}(\theta,Y)$ on 
$\, {\cal M}^{\prime}$ in the form

\begin{equation}
\label{finxpoln}
\varphi^{i}(\theta,X) \,\,\, = \,\,\,
\Phi^{i}_{(in)}\left(\theta + \theta^{*}_{0}(X) +
\theta_{0}(X_{0}) + {1 \over \epsilon}
\int_{X_{0}}^{X} k(J(X^{\prime})) dX^{\prime},
U[J,\theta^{*}_{0}](X)\right)
\end{equation}
and

\begin{equation}
\label{finypoln}
\varphi^{i}(\theta,Y) \,\,\, = \,\,\, 
\Phi^{i}_{(in)}\left(\theta + \theta^{*}_{0}(Y) +
\theta_{0}(X_{0}) + {1 \over \epsilon}
\int_{X_{0}}^{Y} k(J(Y^{\prime})) dY^{\prime},
U[J,\theta^{*}_{0}](Y)\right)
\end{equation}
respectively.

 It is easy to see that the local part of (\ref{jnujmubr})
gives us the expression

\begin{multline*}
{1 \over (2\pi)^{m}}\int_{0}^{2\pi}\!\!\dots\int_{0}^{2\pi}
d^{m}\theta \,\,
A^{\nu\mu}_{0}\left(\Phi^{i}_{(in)}(\theta + s(X), 
U[J,\theta^{*}_{0}](X)), \, \epsilon 
{\partial \over \partial X}\Phi^{i}_{(in)}(\theta + s(X),
U[J,\theta^{*}_{0}](X)), \dots \right) \times  \\
\times \, \delta (X-Y) \quad  +
\end{multline*}
\begin{equation}
\label{locform}
+  \quad \epsilon \langle A^{\nu\mu}_{1}\rangle (J(X))
\delta^{\prime}(X-Y)  \quad  +  \quad  O(\epsilon^{2})
\end{equation}
in the coordinates $\, J(X)$, $\, \theta^{*}_{0}(X)$ and
$\, \theta_{0}(X_{0})$ on $\, {\cal M}^{\prime}$, where
$\, s(X) \equiv \theta^{*}_{0}(X) + \theta_{0}(X_{0}) +
{1 \over \epsilon}
\int_{X_{0}}^{X} k(J(X^{\prime})) dX^{\prime}$.

 Here we used only the main part of (\ref{finxpoln}) 
and its derivatives in the
second term of (\ref{locform}) and replaced $U^{\nu}(X)$
by $J^{\nu}(X)$ according to (\ref{form1}) in the arguments 
of the averaged functions modulo the higher orders of 
$\epsilon$.

 The non-local part of (\ref{jnujmubr}) gives for
(\ref{qqjjbr}) the following equalities:

$$\int \int dX dY \,\,
{1 \over (2\pi)^{m}}\int_{0}^{2\pi}\!\!\!\!\!\dots\int_{0}^{2\pi}
{1 \over \epsilon} \sum_{k\geq 0} \, 
e_{k} \, \epsilon \, q^{\nu}(X) \,
{\partial F^{\nu}_{(k)}\left(\Phi_{(in)}(\theta + s(X),U(X)),
\dots \right) \over \partial X} \, \times $$
$$ \times \,\,\, \nu (X-Y) \,\, \epsilon \, q^{\mu}(Y) \,\,
{\partial F^{\mu}_{(k)}\left(\Phi_{(in)}(\theta + s(Y),U(Y)),
\dots \right) \over \partial Y} \,\, d^{m}\theta \,\,\, = $$

$$= \,\,\, \int \int dX dY \,\,
{1 \over (2\pi)^{m}}\int_{0}^{2\pi}\!\!\dots\int_{0}^{2\pi}
\sum_{k\geq 0} e_{k} \,\,
{\partial^{2} \left[q^{\nu}(X) \nu (X-Y) q^{\mu}(Y)\right]
\over \partial X \partial Y} \, \times $$
$$\times \,\, F^{\nu}_{(k)}\left(\Phi_{(in)}(\theta + s(X),U(X)),
\dots \right) \, 
F^{\mu}_{(k)}\left(\Phi_{(in)}(\theta + s(Y),U(Y)),
\dots \right)  \, d^{m}\theta \,\,\, = $$

$$= \,\,\, \int \int dX dY \,\, 
{1 \over (2\pi)^{m}}\int_{0}^{2\pi}\!\!\dots\int_{0}^{2\pi}
\sum_{k\geq 0} e_{k} \,\, \epsilon \, 
\Big[ \, q^{\nu}_{X}(X) \, \nu (X-Y) \,  
q^{\mu}_{Y}(Y) \,\,\, + $$
$$+ \,\,\, \big( \, q^{\nu}(X) \, q^{\mu}_{Y}(Y) \, - \, 
q^{\nu}_{X}(X) \, q^{\mu}(Y) \big) \, 
\delta (X-Y) \,\, - \,\, q^{\nu}(X) \, q^{\mu}(Y) 
\, \delta^{\prime}(X-Y) 
\Big] \,\, \times $$
$$\times \,\, 
F^{\nu}_{(k)}\left(\Phi_{(in)}(\theta + s(X),U(X)),
\dots \right) \,
F^{\mu}_{(k)}\left(\Phi_{(in)}(\theta + s(Y),U(Y)),
\dots \right) \, d^{m}\theta \,\,\, = $$

$$= \,\,\, \int \int dX dY \,\, 
{1 \over (2\pi)^{m}}\int_{0}^{2\pi}\!\!\dots\int_{0}^{2\pi}
\sum_{k\geq 0} \,\, e_{k} \, \epsilon \, q^{\nu}_{X}(X) 
\,\, \nu (X-Y) \, 
q^{\mu}_{Y}(Y) \,\, \times $$
$$\times \, F^{\nu}_{(k)}\left(\Phi_{(in)}(\theta + s(X),U(X)),
\dots \right) F^{\mu}_{(k)}\left(\Phi_{(in)}(\theta + s(Y),U(Y)),
\dots \right) d^{m}\theta \,\,\, + $$
$$+ \,\,\, \epsilon \sum_{k\geq 0} \, e_{k} \int
\big( \, q^{\nu}(X) \, q^{\mu}_{X}(X) \, - \, 
q^{\nu}_{X}(X) \, q^{\mu}(X) \, \big) \, 
\langle F^{\nu}_{(k)}F^{\mu}_{(k)} \rangle\left(J(X)\right) \,  
dX \,\,\, - $$
$$- \,\,\, \epsilon \sum_{k\geq 0} e_{k} \int
q^{\nu}(X) \, q^{\mu}_{X}(X) \,\,
\langle F^{\nu}_{(k)}F^{\mu}_{(k)} \rangle\left(J(X)\right) \,
dX \,\,\, - $$
$$- \,\,\, \epsilon \int q^{\nu}(X) \, q^{\mu}(X) \,
{1 \over (2\pi)^{m}}\int_{0}^{2\pi}\!\!\!\!\!\dots\int_{0}^{2\pi}
\sum_{k\geq 0} \, e_{k} \, 
F^{\nu}_{(k)}\left(\Phi_{(in)}(\theta + s(X),U(X)),\dots \right)
\, \times $$
\begin{equation}
\label{vyrazh}
\times \,\, {\partial \over \partial X} \,
F^{\mu}_{(k)}\left(\Phi_{(in)}(\theta + s(X),U(X)),\dots \right)
\, d^{m}\theta \, dX  \quad  +  \quad  O(\epsilon^{2})
\end{equation}
where we used the integration by parts for the generalized 
functions.

 We can see now that in the first term of the expression above
in the both regions $X>Y$ and $X<Y$ we have the convolution
with respect to $X$ and $Y$ of the ``slow'' functions 
$\, q^{\nu}_{X}(X) \, q^{\mu}_{Y}(Y)$ with the rapidly oscillating
expressions

$$\langle 
F^{\nu}_{(k)}\left(\Phi_{(in)}(\theta + 
s(X,\epsilon),J(X)),\dots \right)
F^{\mu}_{(k)}\left(\Phi_{(in)}(\theta + 
s(Y,\epsilon),J(Y)),\dots \right)
\rangle $$
in the main order of $\epsilon$. 
Here $\langle \dots \rangle$ means the 
averaging with respect to phases $\theta^{\alpha}$. Now, since 
small $\Delta X$ and $\Delta Y$ lead to the changes of phase 
equal to
$\, {1 \over \epsilon} k^{\alpha}(J(X)) \Delta X + 
O((\Delta X)^{2})$ and
$\, {1 \over \epsilon} k^{\alpha}(J(Y)) \Delta Y +
O((\Delta Y)^{2})$, it is not very difficult to see that in the
sense of ``generalized'' limit (i.e. in the sense of the
convolutions with the ``slow'' functions of $X$ and $Y$) we can
replace these oscillating expressions in the main order of
$\epsilon$ just by their mean values

$$\sum_{k\geq 0} \, e_{k} \,
\langle F^{\nu}_{(k)}\rangle\left(J(X)\right) \,  
\langle F^{\mu}_{(k)}\rangle\left(J(Y)\right) $$
where $\langle \dots \rangle$ means the averaging on the space 
of $m$-phase solutions.

 As for the last term of (\ref{vyrazh}), we recall that its sum 
with the expression arising from the first term 
of the local part in (\ref{locform})

$$\int dX \, q^{\nu}(X) \, q^{\mu}(X) \,
{1 \over (2\pi)^{m}}\int_{0}^{2\pi}\!\!\dots\int_{0}^{2\pi}
d^{m}\theta \,\, \Big[ \,
A^{\nu\mu}_{0}\left(\!\Phi_{(in)}(\theta + s(X),U(X)),\dots 
\right) \,\,\, - $$
$$- \,\,\, \sum_{k\geq 0} \, e_{k} \,
F^{\nu}_{(k)}\!\left(\!\Phi_{(in)}(\theta + s(X),U(X)),\dots 
\right) \,
\epsilon {\partial \over \partial X} \, 
F^{\mu}_{(k)}\!\left(\Phi_{(in)}(\theta + s(X),U(X)),\dots 
\right) \Big] $$
is equal according to (\ref{vazhno}) to

$$\epsilon \int \left(
{\partial \langle Q^{\nu\mu}\rangle\left(J(X)\right) \over
\partial X} \, - \,
\sum_{k\geq 0} e_{k}
{\partial \langle F^{\nu}_{(k)} F^{\mu}_{(k)}
\rangle\left(J(X)\right) \over \partial X} \right)
q^{\nu}(X) \, q^{\mu}(X) \, dX $$
in the leading order of $\epsilon$.

 So, we can write now:

$$\{\int q^{\nu}(X) J^{\nu}(X) dX, 
\int q^{\mu}(Y) J^{\mu}(Y) dY\}|_{{\cal M}^{\prime}} 
\,\,\, = $$
$$= \,\,\, \epsilon \int \left[ q^{\nu}(X) \, q^{\mu}_{X}(X)
\langle A^{\nu\mu}_{1}\rangle\left(J(X)\right) \, - \, 
q^{\nu}_{X}(X) \, q^{\mu}(X) \sum_{k\geq 0} e_{k} 
\langle F^{\nu}_{(k)} F^{\mu}_{(k)}\rangle\left(J(X)\right)  
\right. \,\,\, + $$
$$+ \,\,\, \left. q^{\nu}(X) \, q^{\mu}(X) \, \partial_{X}
\left( \langle Q^{\nu\mu}\rangle\left(J(X)\right) -
\sum_{k\geq 0} e_{k}
\langle F^{\nu}_{(k)} F^{\mu}_{(k)}\rangle\left(J(X)\right)
\right) \right] dX \,\,\, + $$
$$+ \,\,\, \epsilon \int\!\! \int\! \sum_{k\geq 0} \, e_{k} \,
q^{\nu}_{X}(X) \,
\langle F^{\nu}_{(k)}\rangle\left(J(X)\right) \,\, \nu (X-Y) \,
\langle F^{\mu}_{(k)}\rangle\left(J(Y)\right) \,
q^{\mu}_{Y}(Y) \, dX dY \,\,\, + \,\,\, O(\epsilon^{2}) $$ 
 
 After the integration by parts (in the sense of generalized
functions) we can write this expression in the following
``canonical'' form:

$$\{\int q^{\nu}(X) J^{\nu}(X) dX,
\int q^{\mu}(Y) J^{\mu}(Y) dY\}|_{{\cal M}^{\prime}} 
\,\,\, = $$
$$= \,\,\, \epsilon\!\! \int\!\!
\left(\!\langle A^{\nu\mu}_{1}\rangle(J(X)) \, + \,
\sum_{k\geq 0} e_{k}\!\left(
\langle F^{\nu}_{(k)} F^{\mu}_{(k)}\rangle(J(X))\!
\, - \, \langle F^{\nu}_{(k)}\rangle(J(X))
\langle F^{\mu}_{(k)}\rangle(J(X)) \right)
\right) q^{\nu}(X) \, q^{\mu}_{X}(X) \, dX \,\,\, + $$
$$+ \,\,\, \epsilon \int \left(
{\partial \langle Q^{\nu\mu}\rangle(J(X)) \over
\partial X} \, - \, \sum_{k\geq 0} e_{k}
{\partial \langle F^{\nu}_{(k)}\rangle(J(X))
\over \partial X} \,
\langle F^{\mu}_{(k)}\rangle(J(X)) \right) 
q^{\nu}(X) \, q^{\mu}(X) \, dX \,\,\, + $$
$$+ \,\,\, \epsilon\!\! \int \int
\sum_{k\geq 0} \, e_{k} \, q^{\nu}(X) \,\, 
{\partial \langle F^{\nu}_{(k)}\rangle(J(X))
\over \partial X} \,\, \nu (X-Y) \, 
{\partial \langle F^{\mu}_{(k)}\rangle(J(Y))
\over \partial Y} \,\, q^{\mu}(Y) \, dX dY \,\,\, + \,\,\, 
O(\epsilon^{2}) $$
which corresponds to the bracket

$$ \{J^{\nu}(X), J^{\mu}(Y)\}|_{{\cal M}^{\prime}} 
\,\,\, = $$
$$ = \,\,\, \epsilon 
\left(\langle A^{\nu\mu}_{1}\rangle (J(X)) \, + \,
\sum_{k\geq 0} e_{k} \left(
\langle F^{\nu}_{(k)} F^{\mu}_{(k)}\rangle
- \langle F^{\nu}_{(k)}\rangle
\langle F^{\mu}_{(k)}\rangle \right)(J(X)) \right)
\delta^{\prime}(X-Y) \,\,\, + $$
$$ + \,\,\, \epsilon \left(
{\partial \langle Q^{\nu\mu}\rangle (J(X)) \over
\partial X} \, - \, \sum_{k\geq 0} e_{k}
{\partial \langle F^{\nu}_{(k)}\rangle (J(X))
\over \partial X}
\langle F^{\mu}_{(k)}\rangle (J(X)) \right)
\delta (X-Y) \,\,\, + $$
\begin{equation}
\label{jjogrbr}
+ \,\,\, \epsilon \sum_{k\geq 0} \, e_{k} \,
{\partial \langle F^{\nu}_{(k)}\rangle (J(X))
\over \partial X}
\,\, \nu (X-Y) \,
{\partial \langle F^{\mu}_{(k)}\rangle (J(Y))
\over \partial Y} \,\,\, + \,\,\, O(\epsilon^{2}) 
\end{equation}
for the functionals $\, J^{\nu}(X)$.

 So, according to Lemma 2.4 and the remarks above we obtain 
for the Dirac restriction on ${\cal M}^{\prime}$

\begin{equation}
\label{oepsbr}
\{\theta^{*\alpha}_{0}(X), J^{\mu}(Y)\}_{D} \,\,\, = \,\,\, 
O(\epsilon)
\end{equation}
and the relations (\ref{jjogrbr}) for the brackets
$\{J^{\nu}(X), J^{\mu}(Y)\}_{D}$ in the coordinates
$J(X)$, $\theta^{*}_{0}(X)$ and $\theta_{0}(X_{0})$.

 It is evident also that the Dirac brackets
$\{J^{\nu}(X), J^{\mu}(Y)\}_{D}$ on ${\cal M}^{\prime}$ do not
depend in any order of $\epsilon$ on the common initial phase
$\theta_{0}(X_{0})$ because of the invariance of $J^{\nu}(X)$, 
the bracket (\ref{newbr}) and the 
submanifold ${\cal M}^{\prime}$
with respect to the common shifts of $\theta^{\alpha}$.

 The dependence of $\{J^{\nu}(X), J^{\mu}(Y)\}_{D}$ on
$J(X)$, $J_{X}(X)$, $\theta^{*}_{0X}(X), \dots$ is regular at
$\epsilon \rightarrow 0$ and, as can be easily seen from
(\ref{jjogrbr}), we have no dependence of $\theta^{*}_{0}$ 
in the first order of $\epsilon$.
 
 So, it is easy to see now that the Jacobi identities for the
bracket $\{\dots,\dots\}_{D}$ on ${\cal M}^{\prime}$ with 
coordinates $J(X)$, $\theta^{*}_{0}(X)$ and 
$\theta_{0}(X_{0})$, written for the fields
$J^{\nu}(X)$, $J^{\mu}(Y)$ and $J^{\lambda}(Z)$ in the order
of $\epsilon^{2}$, coincide with the corresponding Jacobi
identities for the bracket (\ref{usrbr}).

 So we proved the Jacobi identity for the bracket 
(\ref{usrbr}).

 The skew-symmetry of the bracket (\ref{usrbr}) is just a
trivial corollary of the skew-symmetry of (\ref{newbr}).

\vspace{0.5cm}

 We now prove the invariance of the bracket (\ref{usrbr})
with respect to the choice of the integrals $I^{\nu}$. The
proof is just the same as in the local case and we will just
reproduce it here.

 Under the condition (IV) we have the unique restriction 
of the Poisson bracket (\ref{newbr}) on the submanifold
${\cal M}^{\prime}$ with the coordinates
$J(X)$, $\theta^{*}_{0}(X)$, $\theta_{0}(X_{0})$ in the form
of formal series at $\epsilon \rightarrow 0$.

 So, the two restrictions of (\ref{newbr}), obtained in the
coordinates
$$(J^{\nu}(X), \, \theta^{*\alpha}_{0}(X), \,
\theta^{\alpha}_{0}(X_{0}))$$
and 
$$({\tilde J}^{\nu}(X), \, {\tilde \theta}^{*\alpha}_{0}(X),
\, \theta^{\alpha}_{0}(X_{0}))  \quad , $$
corresponding to the sets $\{I^{\nu}\}$ and
$\{{\tilde I}^{\nu}\}$ (satisfying (A)-(C)) respectively,
should transform one into another after the corresponding
transformation of coordinates

$${\tilde J}^{\nu}(X) \,\,\, = \,\,\,  
{\tilde j}^{\nu}_{(0)}(J(X)) \,\,\, + \,\,\, 
\sum_{k\geq 1} \, \epsilon^{k} \, {\tilde j}^{\nu}_{(k)}
[J,\theta^{*}_{0}](X) $$

$${\tilde \theta}^{*\alpha}_{0}(X) \,\,\, = \,\,\,  
{\tilde \theta}^{*\alpha}_{0}[J,\theta^{*}_{0},\epsilon](X) $$
on ${\cal M}^{\prime}$.

 Now we note that the leading term of (\ref{jjogrbr}), 
coinciding with the bracket (\ref{usrbr}), transforms 
according to the transformation

$${\tilde J}^{\nu}(X) \,\,\, = \,\,\, 
{\tilde j}^{\nu}_{(0)}(J(X))  \quad , $$
which corresponds to the substitution 
$\, {\tilde U}^{\nu}(X) = {\tilde U}^{\nu}(U(X))$ on 
$\, {\cal M}^{\prime}$ view the relation (\ref{form1}).
So, we obtain the second part of the theorem.

{\hfill {\it Theorem 3.1 is proved.}}

\vspace{0.5cm}

{\it Remark.}

 From Theorem 3.1 it also follows in particular that the
procedure (\ref{usrbr}) is insensitive to the
addition of the total derivatives with respect to $x$
to the densities
${\cal P}^{\nu}(\varphi,\varphi_{x},\dots)$. This fact, 
however, can be also obtained from an elementary 
consideration of the definition of bracket (\ref{usrbr}).

\vspace{0.5cm}

{\bf Theorem 3.2}

{\it
 The Hamiltonian functions

$${\bar H}^{\nu} \,\,\, = \,\,\, \int U^{\nu}(X) dX $$
and

$${\bar H} \,\,\, = \,\,\, 
\int \langle {\cal P}_{H}\rangle (U(X)) dX $$
generate view (\ref{usrbr}) local commuting flows, which give
us the Whitham equations for the systems (\ref{nuflows}) and
(\ref{locsys}) respectively. }

\vspace{0.5cm}

{\it Proof.}

 It is easy to check by direct substitution that any of
${\bar H}^{\mu}$ generates the ``conservative'' form

$$U^{\nu}_{T} \, = \, \partial_{X}
\langle Q^{\nu\mu}\rangle (U) $$
of the Whitham's system for the corresponding flow
(\ref{nuflows}). It is easy to see also that this flow 
conserves any of ${\bar H}^{\nu}$ and so all 
${\bar H}^{\nu}$ and ${\bar H}^{\mu}$ commute view the bracket
(\ref{usrbr}). The same property for the Hamiltonian function 
${\bar H}$ (and also for the integral of the averaged density 
of any local integral $I$, commuting with $H$ and $I^{\nu}$
and generating a local flow view (\ref{canform}))
can be now obtained from the invariance of (\ref{usrbr}) with
respect to the set $\{I^{\nu}\}$, since we can use the Hamiltonian
function $H$ in the form (\ref{hamilt}) as one of the 
integrals instead of any of $I^{\nu}$.

{\hfill {\it Theorem 3.2 is proved.}}

\vspace{5mm}

 We can also see that the functionals ${\bar H}^{\nu}$ give us 
conservation laws for our Whitham system.

\vspace{0.5cm}

 From the Theorem 1.1 it follows also that the flows

$$U^{\nu}_{T} \, = \, 
\partial_{X} \langle F^{\nu}_{(k)}\rangle (U) $$
commute with all the local flows, generated by local functionals
$\int h(U) dX$ in the Hamiltonian structure (\ref{usrbr}),
and it can be also seen that they give us the Whitham's 
equations for the corresponding flows (\ref{sflows}).

 It can be easily seen also that the described procedure
can be applied in the same way to the brackets 
(\ref{nonlocbr}) written also in the ``irreducible''
form and not only in the ``canonical'' one.

\vspace{0.5cm}

 The author is grateful to B.A.Dubrovin,
S.P.Novikov, V.L.Alekseev, O.I.Mokhov, M.V.Pavlov 
and E.V.Ferapontov for fruitful discussions.

 The work was supported in part by INTAS (grant INTAS 96-0770)
and RFBR (grant 97-01-00281).

\end{document}